\documentclass[prd,nofootinbib,onecolumn,superscriptaddress]{revtex4}

\usepackage{amsmath}
\usepackage{amssymb}
\usepackage{amsfonts}
\usepackage{amssymb}
\usepackage{dcolumn}
\usepackage{bm}
\usepackage{bbm}
\usepackage{graphicx}
\usepackage{xcolor}
\usepackage{array}
\usepackage{subfigure}
\usepackage{wasysym}
\usepackage{tikz}
\usepackage{hyperref}
\usepackage[normalem]{ulem}

\hypersetup{colorlinks=true,
	breaklinks=true,
	pdfstartview=Fit,
	linkcolor=blue,
	citecolor=blue,
	urlcolor=blue}

\definecolor{lime}{HTML}{A6CE39}
\DeclareRobustCommand{\orcidicon}{
	\begin{tikzpicture}
	\draw[lime, fill=lime] (0,0) 
	circle [radius=0.16] 
	node[white] {{\fontfamily{qag}\selectfont \tiny ID}};
	\draw[white, fill=white] (-0.0625,0.095) 
	circle [radius=0.007];
	\end{tikzpicture}
	\hspace{-2mm}
}

\foreach \x in {A, ..., Z}{
	\expandafter\xdef\csname orcid\x\endcsname{\noexpand\href{https://orcid.org/\csname orcidauthor\x\endcsname}{\noexpand\orcidicon}}
}

\newcommand{\be}{\begin{equation}}
\newcommand{\ee}{\end{equation}}
\def\bl#1\el{\begin{align}#1\end{align}}
\newcommand{\bea}{\begin{eqnarray}}
\newcommand{\eea}{\end{eqnarray}}
\def\l{\left}
\def\r{\right}
\def\ddd{\mathrm{d}}
\def\nn{\nonumber}

\begin{document}

\title{Enhanced primordial gravitational waves from a stiff \\
postinflationary era due to an oscillating inflaton}

\author{Chao Chen \orcidD{}}
\email{iascchao@ust.hk}
\affiliation{Jockey Club Institute for Advanced Study, The Hong Kong University of Science and Technology, Hong Kong, China}

\author{Konstantinos Dimopoulos\orcidB{}}
\email{k.dimopoulos1@lancaster.ac.uk}
\affiliation{Consortium for Fundamental Physics, Physics Department, Lancaster University, Lancaster LA1 4YB, United Kingdom}

\author{Cem Eröncel\orcidC{}}
\email{cem.eroncel@itu.edu.tr}
\affiliation{Istanbul Technical University, Department of Physics, 34469 Maslak, Istanbul, Türkiye}
\affiliation{Faculty of Engineering, MEF University, 34396, Sarıyer, Istanbul, Türkiye}

\author{Anish Ghoshal\orcidA{}}
\email{anish.ghoshal@fuw.edu.pl}
\affiliation{Institute of Theoretical Physics, Faculty of Physics, University of Warsaw, ulica Pasteura 5, 02-093 Warsaw, Poland}

\begin{abstract}

We investigate two classes of inflationary models, which lead to a stiff period after inflation that boosts the signal of primordial gravitational waves (GWs). In both families of models studied, we consider an oscillating scalar condensate, which when far away from the minimum is overdamped by a warped kinetic term, {\it $\acute{a}$} la $\alpha$-attractors. This leads to successful inflation. 
The oscillating condensate is in danger of becoming fragmented by resonant effects when nonlinearities take over. Consequently, the stiff phase cannot be prolonged enough to enhance primordial GWs at frequencies observable in the near future for low orders of the envisaged scalar potential.
However, this is not the case for a higher-order scalar potential. Indeed, we show that this case results in a boosted GW spectrum that overlaps with future observations without generating too much GW radiation to destabilise big bang nucleosynthesis. For example,  taking $\alpha={\cal O}(1)$, we find that the GW signal can be safely enhanced up to $\Omega_{\rm GW}(f)\sim 10^{-11}$ at frequency $f\sim 10^2\,$Hz, which will be observable by the Einstein Telescope.
Our mechanism ends up with a characteristic GW spectrum, which if observed, can lead to the determination of the inflation energy scale, the reheating temperature and the shape (steepness) of the scalar potential around the minimum.
\end{abstract}

\maketitle

\section{Introduction}

The cosmic inflation paradigm, which resolves the horizon and flatness problems and seeds the initial density perturbations for large-scale structure formation~\cite{Brout:1977ix,Sato:1980yn,Guth:1980zm,Linde:1981mu,Starobinsky:1982ee}, also predicts tiny anisotropies in the cosmic microwave background (CMB) measurements \cite{Planck:2018vyg}. After the latest CMB observations, the content of our present Universe in dark matter, dark energy and radiation is now well established in what is known as the $\Lambda$ cold dark matter ($\Lambda$CDM) model. In addition, the improving measurements of the scalar perturbation modes, together with the most recent limits on the presence of tensor modes in the CMB, help narrow down the class of inflation models.
Nevertheless, the history of the Universe from the end of cosmic inflation to the hot big bang phase remains up to now free of any observational constraints. As a consequence, the way the metric perturbation modes evolve after their production during inflation, until the present time, is partly unknown. The consequences of this blackout regarding our Universe history is twofold: (i) We are unable to predict with certainty the energy scale of inflation and (ii) the number of $e$-folds of cosmic inflation, which is essential to constrain cosmic inflation models from the CMB measurement, is not precisely determined.

In the vanilla $\Lambda$CDM model, it is frequently assumed that the cosmic inflation era is followed immediately by the radiation-dominated era of the hot big bang phase of the cosmological history. 
Since the slow-roll inflation is expected to produce a nearly scale-invariant spectrum of linear tensor perturbations that is relatively feeble as compared to the sensitivity of present and near-future gravitational wave (GW) detectors, it is expected that a Universe exclusively dominated by radiation and matter after inflation would not lead to any measurable primordial GW signal in the near future. However, we would like to highlight that the Universe can only become radiation dominated at the end of inflation under very restrictive assumptions. Indeed, to release all of its energy density right after it exits the phase of slow roll, the inflaton must decay immediately into ordinary radiation. Such a fast decay of the inflaton field requires the existence of large interaction terms between the inflaton field and Standard Model (SM) fields.

However, sizable interactions of the inflationary sector with the SM are not motivated by any strong theoretical argument. Additionally, they were also shown to substantially affect the inflationary dynamics \cite{Buchmuller:2014pla,Buchmuller:2015oma,Argurio:2017joe,Heurtier:2019eou} or the stability of the SM Higgs boson \cite{Enqvist:2016mqj, Kost:2021rbi}.  
Moreover, there is a danger that significant interaction terms may spoil the flatness of the inflaton potential.
Furthermore, in order to decay efficiently after inflation ends, the inflaton field also needs to oscillate around the minimum of its potential, such that its coherent oscillations quickly get damped through SM particle production. This relies on the idea that the inflation potential minimum stands relatively close in field space from the point where inflation ends. However, numerous runaway scalar potentials can be used to realize cosmic inflation, which do not have a finite minimum or whose minimum is very far away from the location in field space where inflation ends. 

This is, for instance, the case of quintessential inflationary scenarios  \cite{Peebles:1998qn, Dimopoulos:2001ix, Akrami:2017cir, Bettoni:2021qfs} or, more generally, nonoscillatory inflation models \cite{Ellis:2020krl}. In these models the inflaton keeps rolling along its potential for a quite long period of time time after inflation ends. In such cases, the production of SM particles is more difficult to achieve, however, can be realized through gravitational particle production~\cite{Ford:1986sy,Chun:2009yu} or other reheating mechanisms, e.g., instant preheating \cite{Felder:1998vq,Dimopoulos:2017tud}, curvaton reheating \cite{Feng:2002nb,BuenoSanchez:2007jxm}, and Ricci reheating \cite{Dimopoulos:2018wfg,Opferkuch:2019zbd,Bettoni:2021zhq,Laverda:2023uqv,Figueroa:2024asq} to cite few examples.\footnote{See also reheating by evaporation of primordial black holes \cite{Dalianis:2021dbs}, warm quintessential inflation \cite{Dimopoulos:2019gpz,Rosa:2019jci} or the large-scale isocurvature perturbations~\cite{Jiang:2018uce}.} The inflation sector thus only transfers at most a fraction of its energy density when SM particles are produced. The Universe therefore undergoes a phase of {\em kination} \cite{Joyce:1997fc,Gouttenoire:2021jhk}, where the kinetic energy of the inflaton scalar field is the main source of energy in the Universe and decreases quickly with expansion as $\rho_\phi\propto a^{-6}$ before radiation starts dominating and the hot big bang phase starts. The corresponding barotropic parameter during kination is $w=1$, stiffer than the barotropic parameter during radiation domination (RD) ($w=1/3$) or matter domination (MD) ($w=0$).

The tensor perturbation modes that reenter the horizon after inflation during kination are not characterized by a flat spectrum, as with the modes reentering the horizon during RD. Instead, for frequencies that correspond to the period of kination, the GW spectrum features a peak, which is larger the longer kination lasts. This boosted spectrum peaks at the highest frequency possible, which corresponds to the end of inflation, when kination begins. Such frequencies are beyond observational capability in the near future. However, kination cannot be extended down to frequencies low enough to overlap with future GW surveys, because the peak in the GW spectrum would be too large, as their energy density would destabilize the delicate process of big bang nucleosythesis (BBN) \cite{Giovannini:1999bh,Sahni:2001qp,Dimopoulos:2002hm}.

One way of boosting the GW signal down to observable frequencies without disturbing BBN  is considering that the stiff phase following inflation is not as stiff as kination proper but it is a period when the barotropic parameter of the Universe lies in the range \mbox{$1/3<w<1$}~\cite{Tashiro:2003qp, Bernal:2020ywq,Ghoshal:2022ruy,Berbig:2023yyy,Barman:2023ktz,Mishra:2021wkm,Bernal:2019lpc}.\footnote{Other ways have also been put forward, for example in Ref.~\cite{SanchezLopez:2023ixx}, the GW peak is truncated by considering that inflation is followed first by a period called hyperkination before kination proper. In Refs.~\cite{Cai:2020ovp, Cai:2023ykr}, GWs can be boosted within a narrow band through the parametric resonance.}
Indeed, recently in Ref.~\cite{Figueroa:2019paj}, it was argued that, to make contact with the forthcoming Laser Interferometer Space Antenna (LISA) observations, the stiff period after inflation must be in the range \mbox{$0.46\lesssim w\lesssim 0.56$} with a high inflationary scale $H_{\rm inf} \sim 10^{13}$ GeV and the reheating temperature in the range \mbox{1 MeV$\,\lesssim T_{\rm reh}\lesssim\,$150 MeV}. A model realization of this possibility was presented in Ref.~\cite{Dimopoulos:2022mce}, where \mbox{$w\approx1/2$} was considered.

In this paper we will consider the possibility that the end of inflation does not continue right away to the hot big bang phase, but instead is followed by a phase featuring a stiff equation of state which is not due to the field rolling down a runaway potential, as in Ref.~\cite{Dimopoulos:2022mce}, but oscillating instead in a $2n$th-order monomial potential \mbox{$V\propto\varphi^{2n}$}. We study two possibilities that may give rise to such a potential. In one, we consider a field with a scalar potential that is truly monomial, motivated by a variety of models, based on fundamental theory; see for example, Refs.~\cite{Lyth:1995ka,Lazarides:1985ja,Lazarides:1986rt}.
In the other, we consider a quasiharmonic periodic potential, that may correspond to an axion-like particle, possibly in the context of the string axiverse \cite{Arvanitaki:2009fg}. In both cases, the scalar field is also characterized by a noncanonical kinetic term, following the $\alpha$-attractors idea \cite{Kallosh:2013yoa,Kallosh:2022ggf}, such that before engaging in oscillations, it successfully drives a period of inflation. We show that our setup can naturally lead to \mbox{$1/3<w<1$}. The hope is that, considering an oscillating inflaton field, we may manage to generate an observable, characteristic peak in the GW spectrum, without the need of substantial tuning. Such peaked feature will help us determine the physical information (including the inflationary energy scale and potential's shape, as well as the reheating temperature) regarding the early Universe via the near-future GW experiments.

The paper is organized as follows: In Secs.~\ref{sec:monomial-model} and \ref{sec:sinusoidal-model} we discuss the monomial model, which we call T-model due to the existence of the noncanonical kinetic term and the harmonic model, respectively. Afterward, we discuss preheating phenomenology in Sec.~\ref{sec:frag} and generation and propagation of primordial gravitational waves in Sec.~\ref{sec:gw}. Finally, we end with discussing the key aspects of our analysis and conclusions in Sec.~\ref{sec:conclusion}.
We use natural units, where \mbox{$c=\hbar=k_{\rm B}=1$} and \mbox{$8\pi G=m_P^{-2}$}, with \mbox{$m_P=2.43\times 10^{18}\,$GeV} being the reduced Planck mass.

\section{T-Model inflation}
\label{sec:monomial-model}

The idea is to consider a simple $2n$th-order monomial scalar potential. This is understood as a perturbative expansion of the scalar potential around the vacuum expectation value (VEV) (taken at zero) of a scalar field $\varphi$, with a simplifying $Z_2$ symmetry. 
An example is a flaton field \cite{Lyth:1995ka}, where $V\propto\varphi^6$ with the quartic self-interaction term being absent, while the quadratic mass term is discussed later on.
Other examples are supersymmetric flat directions, such as discussed in Refs.~\cite{Lazarides:1985ja, Lazarides:1986rt}, where $V\propto\varphi^6$ and $V\propto\varphi^{10}$, respectively.
We also consider that our scalar field is characterized by a noncanonical kinetic term, which features two poles around the VEV, following the $\alpha$-attractors construction \cite{Kallosh:2013yoa,Kallosh:2022ggf} due to the geometry in field space, e.g., characterized by a nontrivial K\"{a}hler metric. As we demonstrate, this construction successfully generates the inflationary plateau for the canonically normalized scalar field $\phi$. 
After inflation, the $\phi$ field exits the flat region of the potential and $\varphi$ becomes effectively canonically normalized. As a result, it oscillates around its VEV (i.e., zero) in a $2n$th-order potential having an average barotropic parameter of \mbox{$w\approx (n-1)/(n+1)$}, which results in a stiff phase that would produce observable gravitational waves if reheating were appropriately inefficient~\cite{Figueroa:2019paj}.

\subsection{Inflationary dynamics}

The Lagrangian density of the model we consider is simply
\begin{equation}
    {\cal L}=-\frac{\frac12(\partial\varphi)^2}{(1-\varphi^2/M^2)^2}
    - \frac{1}{(2n)!}\lambda\frac{\varphi^{2n}}{m_P^{2n-4}}\,,
    \label{Vvarphi6}
\end{equation}
where \mbox{$(\partial\varphi)^2\equiv\partial_\mu\varphi\,\partial^\mu\varphi$} and \mbox{$n>2$}.
In order to switch to the canonical field $\phi$, we employ the transformation:
\begin{equation}
    \varphi=M\tanh(\phi/M)\,,
    \label{phivarphi}
\end{equation}
and the potential becomes
\begin{equation}
    V(\phi)=\frac{1}{(2n)!}
    \lambda
    \frac{M^{2n}}{m_P^{2n-4}}
    \tanh^{2n}(\phi/M)\,.
    \label{V6}
\end{equation}
This is the well-known and researched T-model inflation \cite{Carrasco:2015pla,Carrasco:2015rva}.\footnote{For a complete analysis of such models
in the context of dynamical systems, see Ref.~\cite{Alho:2017opd}.}

For the derivatives, we find
\begin{equation}
    V'(\phi)=
    \frac{\lambda}{(2n-1)!}
    \frac{M^{2n-1}}{m_P^{2n-4}}
    \frac{\tanh^{2n-1}(\phi/M)}{\cosh^2(\phi/M)}
    \label{V'6}
\end{equation}
and
\begin{eqnarray}
    V''(\phi) & = &
    \frac{\lambda}{(2n-1)!}
    \frac{M^{2n-2}}{m_P^{2n-4}}
    \frac{\tanh^{2n-2}(\phi/M)
    }{\cosh^4(\phi/M)}\nonumber
    \\
    & & \times
    \left[
    (2n+1)-2\cosh^2(\phi/M)
    \right]\,,
    \label{V"6}
\end{eqnarray}
where the prime denotes derivative with respect to the canonical field $\phi$.
We then find the slow-roll parameters as
\begin{widetext}
\begin{equation}
    \varepsilon\equiv\frac12 m_P^2\l(\frac{V'}{V}\r)^2=2n^2
    \frac{m_P^2}{M^2}\frac{1}{[\cosh^2(\phi/M)-1]\cosh^2(\phi/M)}
    =8n^2
    {m_P^2 \over M^2} { 1 \over \sinh^2(2\phi/M)}
    \label{eps6}
\end{equation}
\end{widetext}
and 
\begin{widetext}
\begin{equation}
    \eta\equiv m_P^2\frac{V''}{V}=2n
    \frac{m_P^2}{M^2}\frac{
    (2n+1)-2\cosh^2(\phi/M)}{[\cosh^2(\phi/M)-1]\cosh^2(\phi/M)}
    = 8n
    {m_P^2 \over M^2} {2n
    - \cosh(2\phi/M) \over \sinh^2(2\phi/M)} \,.
    \label{eta6}
\end{equation}
\end{widetext}
The spectral index of the scalar curvature perturbation is
\begin{widetext}
\begin{equation}
    n_s=1-6\varepsilon+2\eta=
    1-16n
    \frac{m_P^2}{M^2}
    \frac{ n+\cosh(2\phi/M)}{\sinh^2(2\phi/M)}=
    1-\frac{2}{n}
    \left[(n-1)+2\cosh^2(\phi/M)\right]\,\varepsilon\,.
    \label{nsphi}
\end{equation}
\end{widetext}
For the number of $e$-folds until the end of inflation, we find
\begin{equation}
    N =\frac{1}{m_P^2}\int_{\phi_{\rm end}}^{\phi}\frac{V{\rm d}\phi}{V'}
    =\frac{1}{8n}
    \frac{M^2}{m_P^2}
    \l[ \cosh(2\phi/M)-\cosh(2\phi_{\rm end}/M)\r]\,,
    \label{N6}
\end{equation}
where ``end'' denotes the end of inflation.
Demanding that \mbox{$\varepsilon(\phi_{\rm end})=1$} we can estimate that
\begin{equation}
    \cosh(2\phi_{\rm end}/M)=
    \sqrt{1+8n^2
    \frac{m_P^2}{M^2}}
\,.
\label{phiend6}    
\end{equation}
The above implies that
\begin{equation}
    \cosh\l(2\phi(N)/M\r)
    = \sqrt{1+8n^2 \frac{m_P^2}{M^2}}
    +8n \frac{m_P^2}{M^2}N\,.
\label{phiN6}
\end{equation}
Employing this in Eq.~\eqref{eps6}, we obtain
\begin{equation}
    \varepsilon=8n^2
    \frac{m_P^2}{M^2}\frac{1}{\l(\sqrt{1+8n^2
    \frac{m_P^2}{M^2}}+8n
    \frac{m_P^2}{M^2}\,N\r)^2-1}\,.
    \label{eps6N}
\end{equation}
Similarly, Eq.~\eqref{nsphi} becomes
\begin{equation}
    n_s=1-16n
    \frac{m_P^2}{M^2}\frac{\sqrt{1+8n^2
    \frac{m_P^2}{M^2}}+8n
    \frac{m_P^2}{M^2}\,N+
    n}{\l(\sqrt{1+8n^2
    \frac{m_P^2}{M^2}}+8n
    \frac{m_P^2}{M^2}\,N\r)^2-1}\,,
    \label{nsN}
\end{equation}
where we also used Eq.~\eqref{phiN6}.

For the tensor-to-scalar ratio we find
\begin{equation}
    r=16\varepsilon=\frac{
    128n^2\frac{m_P^2}{M^2}}{\l(\sqrt{1+8n^2
    \frac{m_P^2}{M^2}}+8n
    \frac{m_P^2}{M^2}\,N\r)^2-1}\,.
    \label{r6N}
\end{equation}

In the limit \mbox{$\phi\gg\phi_{\rm end}$}, the above reduces to \mbox{$n_s\simeq 1-2/N$} and \mbox{$r\simeq\frac{M^2}{m_P^2}\frac{2}{N^2}$}, which are the usual findings of $\alpha$-attractors. 
Indeed, using the $\alpha$-attractors relation \cite{Kallosh:2013yoa,Kallosh:2022ggf}
\begin{equation}
M\equiv\sqrt{6\alpha}\,m_P\,,
    \label{Malpha}
\end{equation}
we obtain the standard result \mbox{$r\simeq 12\alpha/N^2$}. The observational bound \mbox{$r<0.03$} \cite{Galloni:2022mok,Tristram:2021tvh} suggests that \mbox{$\alpha<9$} for \mbox{$N=60$}. In our case, the stiff period after inflation  increases $N$ somewhat, so the bound is more likely \mbox{$\alpha\lesssim 10$}.
\footnote{The parameter $\alpha$ may have a multitude of values. Some very important examples are the well-known Starobinsky model \cite{Starobinsky:1980te}, the Higgs inflation model ($\alpha=1$) \cite{Bezrukov:2007ep}, and the Goncharov-Linde model ($\alpha=1/9$) \cite{Goncharov:1984jlb,Linde:2014hfa}, and others. Furthermore, other very interesting examples are also related to superstring-inspired scenarios, which suggest \mbox{$3\alpha=1,2,3,4,5,6,7$} \cite{Ferrara:2016fwe,Kallosh:2017ced} (e.g., fibre inflation with $\alpha=2$ and 1/2 \cite{Cicoli:2008gp,Kallosh:2017wku}) or in no-scale supergravity, which accommodates arbitrary values both \mbox{$\alpha<1$} or \mbox{$\alpha>1$} \cite{Ellis:2019bmm,Ellis:2019hps}.}

With $N\gg 1$, it is easy to show that \mbox{$\phi\gg\phi_{\rm end}$} for any
\begin{equation}
   \alpha<  \frac{2n^2}{3}
    \l(-1+\sqrt{1+\frac{4N^2}{n^2}}\r)\simeq\frac{4nN}{3}\,,
\end{equation}
where we considered that \mbox{$n<4N$}. When \mbox{$N\simeq 60$}, this means \mbox{$\alpha < 40\,n$}, which is well satisfied.

The amplitude of primordial curvature perturbation is calculated as
\begin{widetext}
\begin{equation}
    {\cal P}_\zeta=\frac{1}{12\pi^2}\frac{V^3}{m_P^6(V')^2}=
    \frac{\lambda}{48\pi^2n^2(2n)!}
    \l(\frac{M}{m_P}\r)^{2n+2}
    \tanh^{2n+2}(\phi/M)
    \cosh^4(\phi/M)\,.
    \label{Pzphi}
\end{equation}
\end{widetext}
The requirement that
\mbox{$\phi(N)\gg\phi_{\rm end}$} 
results in
\begin{equation}
    \lambda\l(\frac{M}{m_P}\r)^{2n-2}
    \simeq\frac{
    3(2n)!\pi^2{\cal P}_\zeta}{N^2}
    \label{lambda}\,.
\end{equation}
Using Eq.~\eqref{Malpha} and that the Cosmic Background Explorer constraint \mbox{${\cal P}_\zeta=2\times10^{-9}$} and \mbox{$N\simeq 60$}, demanding a perturbative $\lambda<1$, we find the lower bound
\begin{equation}
    \alpha \gtrsim \frac16[10^{-11}(2n)!]^{\frac{1}{n-1}}\,.
    \label{alphabound}
\end{equation}

For the energy scale of inflation we have
\begin{equation}
    V_{\rm inf}^{1/4}
=\l(\frac{\lambda}{(2n)
    !}\r)^{1/4}
    \l(\frac{M}{m_P}\r)^{n/2}
    m_P
    \simeq \l(\frac{M}{m_P}\r)^{1/2}\l(\frac{3\pi^2{\cal P}_\zeta}{N^2}\r)^{1/4} m_P
    \simeq 3\times 10^{-3}\alpha^{1/4}\,m_P\,,
    \label{Vinf}
\end{equation}
where we used Eqs.~\eqref{V6}, \eqref{Malpha} and \eqref{lambda}. 
The above result is independent of the value of $n$ and it suggests that \mbox{$V_{\rm inf}^{1/4} \simeq 7.7\times 10^{15}\,$GeV$\,\times\,\alpha^{1/4}$}, i.e., near the grand unified theory (GUT) scale, as expected.

So inflation seems to work fine because, when \mbox{$N\simeq 60$}, the $\alpha$-attractors are known to produce values of the inflationary observables $n_s$ and $r$ that are in excellent agreement with observations \cite{Kallosh:2013yoa,Kallosh:2022ggf}. 

\subsection{Dynamics during oscillations}

After inflation the field oscillates in a $2n$th-order monomial potential. This means that its barotropic parameter is~\cite{Turner:1983he}
\begin{equation}
    w\approx\frac{n-1}{n+1}
    \,,
    \label{w6}
\end{equation}
where \mbox{$V\propto\phi^{2n}$}. We write ``$\approx$'' instead of ``='' here because there are tiny deviations from the exact equality \cite{Barman:2023ktz}, which, however, are largely negligible and will be ignored hereafter.

We do not have to go into details about the radiation production. It could be due to some other degree of freedom or due to the decay of the $\phi$ condensate itself. In the latter case we need to have the potential supplemented with a quadratic mass term. The reason is that, otherwise, the decay products of the quanta of the oscillating condensate will be able to decay back, meaning the inverse decay would also be possible, so the scalar field would not be able to decay completely. This is so because the density of the oscillating condensate decreases faster than relativistic decay products (\mbox{$\rho_\phi\propto a^{-3(1+w)}=a^{-6n/(n+1)}$}, which is faster than \mbox{$a^{-4}$} for all \mbox{$n>2$}), so the resulting thermal bath will be partly composed of $\varphi$ particles, decaying back and forth. In contrast, if the oscillating condensate is dominated by the quadratic mass term then \mbox{$\rho_\phi\propto a^{-3}$} and the density of the relativistic decay products reduces faster, which means that the reverse interaction, even if it occurs, would create a negligible contribution to $\rho_\phi$. As a result, the condensate can decay fully into a thermal bath that is composed predominantly by its decay products only.

Because, after inflation and the onset of oscillations, we have \mbox{$\varphi\ll M$}, the $\varphi$ field is approximately canonical, i.e. \mbox{$\varphi\simeq\phi$}. Then, we can consider a previously subdominant (negligible during inflation) quadratic term such that the potential is
\begin{equation}
    V(\phi)\simeq\frac12m^2\phi^2+
    \frac{1}{(2n)!}
    \lambda\frac{\phi^{2n}}
    {m_P^{2n-4}}\,.
    \label{Vnew}
\end{equation}
The energy density is \mbox{$\rho_\phi=V(\Phi)$}, where $\Phi$ is the oscillation amplitude, with \mbox{$\Phi=\Phi(t)$}. Equating the quadratic and the $2n$th-order term we find the value $\Phi_{\rm x}$ when the quadratic term becomes important,
\begin{equation}
    \Phi_{\rm x}^{2n-2}=\frac{(2n)!}{2\lambda}\,m^2\,m_P^{2n-4}\,.\label{Phim}
\end{equation}
The corresponding energy density is
\begin{equation}
    \rho_\phi^{\rm x}=\frac12m^2\Phi_{\rm x}^2=
    \left(2^{-n}\,\frac{(2n)!}{\lambda}m^{2n} m_P^{2n-4}\right)^{1/(n-1)}\,.
    \label{rhom}
\end{equation}
In view of Eq.~\eqref{lambda}, the above can be recast as
\begin{equation}
    \rho_{\rm x} \simeq 6\alpha\left(2^{-n}\,\frac{N^2}{3\pi^2{\cal P}_\zeta}m^{2n}m_P^{2n-4}\right)^{1/(n-1)},
    \label{rhom+}
\end{equation}
where we also used Eq.~\eqref{Malpha}. 

After domination from the quadratic term, the energy density of the oscillating inflaton is diluted as nonrelativistic matter \mbox{$\rho\propto a^{-3}$}, as also mentioned above. If the oscillating condensate continued to dominate the Universe, then the effective matter-dominated period would suppress the production of gravitational waves, so this period should not last long. In fact, to maximize the amplitude of the produced GWs, we should demand that the moment that the mass term dominates coincides with the moment that the condensate decays, such that the effective matter-dominated period is eliminated. That is, we must require \mbox{$\rho_\phi^{\rm x}\simeq \rho_{\rm x}\simeq\rho_{\rm reh}$}, where the energy density at reheating is
\begin{equation}
    \rho_{\rm reh}=\frac{\pi^2}{30}g_*T_{\rm reh}^4 \,,
\label{rhoreh}
\end{equation}
where $T_{\rm reh}$ is the reheating temperature and $g_*$ is the number of effective relativistic degrees of freedom. 

From Eqs.~\eqref{rhom+} and \eqref{rhoreh}, we find
\begin{equation}
    m \simeq
    \sqrt 2\left(\frac{\pi^2 g_*}{180\alpha}\right)^{(n-1)/2n}\left(\frac{3\pi^2{\cal P}_\zeta}{N^2}\right)^{1/2n}
m_P^{(2-n)/n}\,T_{\rm reh}^{2(n-1)/n}\,.
    \label{mTreh}
\end{equation}
which can be used to estimate $m$ for a given reheating temperature and a given value of $n$.

Alternatively, we can assume that reheating happens through other means (e.g., Ricci reheating \cite{Dimopoulos:2018wfg,Opferkuch:2019zbd,Bettoni:2021zhq,Laverda:2023uqv,Figueroa:2024asq}, or curvaton reheating \cite{Feng:2002nb,BuenoSanchez:2007jxm}). Then, Eq.~\eqref{mTreh} becomes only upper bound, ensuring that reheating occurs before the quadratic mass term dominates the oscillating condensate, such that there is no effective matter-dominated period.

\section{The Harmonic Model}
\label{sec:sinusoidal-model}

One way to avoid introducing the mass altogether is to consider the non-perturbative potential
\begin{equation}
    V(\varphi)=V_0[1-\cos(\varphi/M)]^n\,,
    \label{VALP}
\end{equation}
in place of the original $V\propto\varphi^{2n}$ potential in Eq.~\eqref{Vvarphi6}. A potential of this form has already been considered in the literature in the context of early dark energy \cite{Poulin:2018cxd} and is possible to justify in the context of the string axiverse \cite{Arvanitaki:2009fg,Kamionkowski:2014zda,Poulin:2018dzj}. The potential in Eq.~\eqref{VALP} reduces to the $2n$th-order potential when $\varphi\ll M$, when $\cos(\varphi/M)\simeq 1-\frac12(\varphi/M)^2$. The previous discussion is the same if we take
\begin{equation}
    V_0=
    2^n\frac{\lambda}{(2n)!}\frac{M^{2n}}{m_P^{2n-4}}=2^n\,\frac{3\pi^2{\cal P}_\zeta}{N^2}(M\,m_P)^2
    \label{V0}\,,
\end{equation}
such that, when \mbox{$\varphi\ll M$} we have \mbox{$V\simeq\frac{1}{(2n)!}\lambda\varphi^{2n}/m_P^{2n-4}$} as before. In the last equation in the above we also took into account Eq.~\eqref{lambda}.

In view of Eq.~\eqref{Malpha}, the above suggests 
\begin{equation}
    V_0\simeq 2^n
    \times 10^{-10}\alpha\;m_P^4\,,
    \label{V0_2}
\end{equation}
which implies that
\mbox{$V_0^{1/4}=2^{n/4} \alpha^{1/4}\times 7.7\times 10^{15}\,$GeV}, that is, $V_0^{1/4}$ is comparable to the energy scale of grand unification, provided $\alpha$ is not extremely small.

For the inflation scale, the plateau exists when we approach the kinetic pole at \mbox{$\varphi\rightarrow M$} (without loss of generality we assume $\varphi>0$). In this case, the potential energy is $V_{\rm inf} \sim \frac{1}{(2n)!}\, \lambda M^{2n}/m_P^{2n-4}$.
The inflationary observables $n_s$ and $r$ must be calculated numerically this time. The good thing is that there is no upper bound on $\lambda$, which is not a perturbative coupling now. Thus, the corresponding lower bound on $\alpha$ in Eq.~\eqref{alphabound} is no more. However, we need that the axion decay constant $M$ must not be super-Planckian. In view of Eq.~\eqref{Malpha}, this implies the requirement \mbox{$\alpha\lesssim 4$}. 

With the choice in Eq.~\eqref{VALP} we can safely ignore $m$ altogether. We do not reheat the Universe through the field decay, but we have no problem with radiative corrections, as presumably the field is some axion-like particle in the context of the string axiverse \cite{Arvanitaki:2009fg,Kamionkowski:2014zda,Poulin:2018dzj}.

\subsection{Inflationary dynamics}

\begin{figure}
    \centering
    \includegraphics[width=0.5\textwidth]{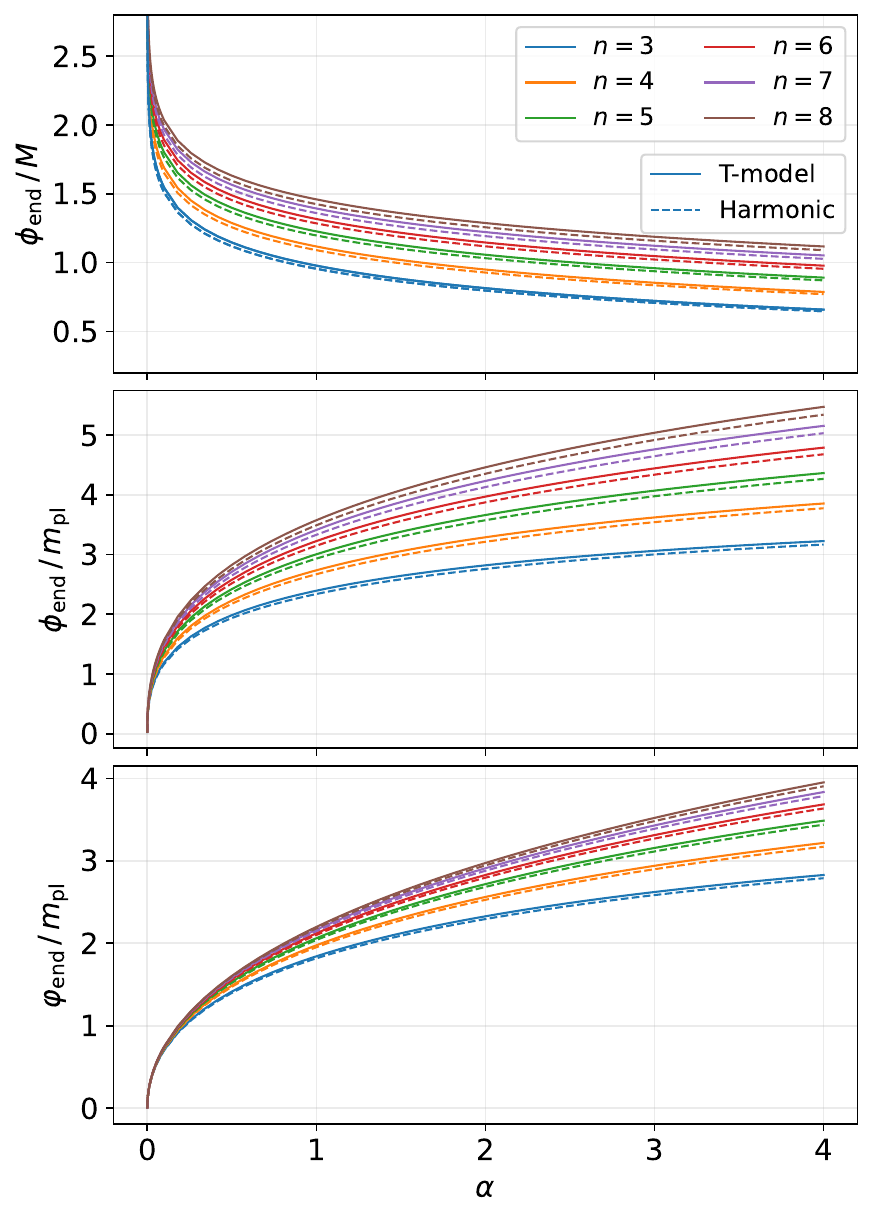}
    \caption{The (non) canonical inflaton field values ($\varphi$) $\phi$ at the end of inflation in terms of $\alpha$, derived by demanding $\varepsilon(\phi_{\rm end})=1$, for the T-model and harmonic potentials, Eqs.~\eqref{Vvarphi6} and \eqref{VALP}, 
    respectively. As shown, the values are Planckian for \mbox{$\alpha={\cal O}(1)$}.
    }
    \label{fig:phiend}
\end{figure}

Following similar calculations, we derive the corresponding results for the harmonic potential \eqref{VALP} as follows:

\bl
V(\phi) &= V_0 \l[1 - \cos \l( \tanh{\phi\over M} \r) \r]^n ~,\\
V'(\phi) & = \frac{n\,V_0}{M}\l[1 - \cos\l( \tanh{\phi\over M} \r) \r]^{n- 1}
\frac{\sin[\tanh(\phi/M)]}{\cosh^2(\phi/M)}~,\\
V''(\phi) &=\frac{n\,V_0}{M^2}\frac{\left\{1-\cos[\tanh(\phi/M)]\right\}^{n-1}}{\cosh^4(\phi/M)}~ \nonumber \\
&\times \Big\lbrace(n-1)\frac{\sin^2[\tanh(\phi/M)]}{1-\cos[\tanh(\phi/M)]}
+\cos[\tanh(\phi/M)] 
-2\sin[\tanh (\phi/M)]\tanh(\phi/M)\cosh^2(\phi/M)\Big\rbrace
\el
The slow-roll parameters are
\bl
\varepsilon = \frac12 m_P^2 \l(\frac{V'}{V}\r)^2
= 
{n^2 m_P^2 {\rm sech}^4(\phi/M)
\sin^2\l[\tanh(\phi/M)
\r] \over 2 M^2 \l\{ 1 - \cos\l[\tanh(\phi/M)
\r] \r\}^2} ~
\el
and
\bl
\eta = m_P^2\frac{V''}{V}=n\left(\frac{m_P}{M}\right)^2 \frac{{\rm sech}^4(\phi/M)}{1-\cos[\tanh(\phi/M)]}\Big\lbrace &(n-1)\frac{\sin^2[\tanh(\phi/M)]}{1-\cos[\tanh(\phi/M)]}+ \cos[\tanh(\phi/M)] \nonumber \\
    &-2\sin[\tanh(\phi/M)]\tanh(\phi/M)\cosh^2(\phi/M)\Big\rbrace ~.
\el
Demanding that $\varepsilon(\phi_{\rm end})=1$, one can estimate $\phi_{\rm end}$ numerically, as shown in Fig. \ref{fig:phiend}.

 The spectral index of the curvature perturbation is
\begin{widetext}
    \bl
    n_s - 1&= 2\eta - 6\varepsilon\nonumber \\
    &=\left(\frac{m_P}{M}\right)^2 \Big\lbrace-n(2+n) 
    \frac{{\rm sech}^4(\phi/M)\sin^2[\tanh(\phi/M)]}{\left\{1-\cos[\tanh(\phi/M)]\right\}^2}
    + 2n \frac{{\rm sech}^4(\phi/M)\cos[\tanh(\phi/M)]}{1-\cos[\tanh(\phi/M)]}\nonumber \\
    &\qquad\qquad\quad\;\;\,
    - \frac{4n\sin[\tanh(\phi/M)]\tanh(\phi/M)\,{\rm sech}^2(\phi/M)}{1-\cos[\tanh(\phi/M)]}\Big\rbrace ~.
    \el
\end{widetext}
We can find the field value $\phi_{\ast}$ at the time when the pivot scale exits the horizon as
\be
N_{\ast} \equiv N(\phi_{\ast}) = \frac{1}{m_P^2}\int_{\phi_{\rm end}}^{\phi_*}
\frac{V(\phi)}{V'(\phi)}\,\ddd{\phi}\,,
\ee
where $k_p=0.05\,\textrm{Mpc}^{-1}$ is the pivot scale and $\phi_{\rm end}$ is determined via $\varepsilon(\phi_{\rm end})=1$.

The curvature power spectrum is calculated as
\begin{widetext}
\be
{\cal P}_\zeta = \frac{1}{12\pi^2}\frac{V^3}{m_P^6(V')^2}
= \frac{V_0/m_P^4}{12n^2\pi^2} \left(\frac{M}{m_P}\right)^2 \frac{\left\{1-\cos[\tanh(\phi/M)]\right\}^{n+2} \cosh^4(\phi/M)}{\sin^2[\tanh(\phi/M)]}~.
\ee
\end{widetext}
At the pivot scale we have \mbox{$A_s=\mathcal{P}_\zeta (k_{p})$}, where \mbox{$\ln(10^{10} A_s)= 3.044\pm 0.014$}.
Note that we are only able to numerically calculate the above quantities and impose the constraints on $V_0$. For example, taking $n=3$, when $\alpha\approx 1$, we determine $\phi_{\rm end}/m_P \simeq 2.34$ from $\varepsilon(\phi_{\rm end}) = 1$. If $N = 60$, then we find $\phi_{\ast}/m_P \simeq 7.48$. Hence, the power spectrum evaluated at $k_p$ is given by
\be\label{V0ex}
A_s\equiv {\cal P}_\zeta(k_p) \simeq \frac{V_0}{m_P^4}\times 2.0
\Rightarrow V_0 \simeq 1.0 \times 10^{-9}\,m_P^4 
\Rightarrow 
V_0^{1/4}=1.4
\times 10^{16}\,{\rm GeV}\,.
\ee
The corresponding inflationary observables are $r \simeq 0.003$ and $n_s = 0.967$.
The above value of $V_0$ in Eq.~\eqref{V0ex} is not too different from the result obtained in the monomial case, given by Eq.~\eqref{V0_2}.

The stiff period is expected to increase $N_{\ast}$; the number of $e$-folds of remaining inflation that correspond to the exit of the cosmological scales from the horizon.
Fixing the reheating temperature $T_{\rm reh}$ allows us to calculate the necessary $N_{\ast}$ via the relation \cite{Drewes:2017fmn,Drewes:2019rxn}
\begin{widetext}
\be
N_{\ast} \simeq 67 - \ln\left(\frac{k_p}{a_0 H_0}\right) + \frac{1}{4}\ln\left(\frac{V^2(\phi_{\ast})}{m_P^4\,\rho_{\rm end}}\right) + \frac{1 - 3 w}{12(1 + w)}\ln\left(\frac{\rho_{\rm reh}}{\rho_{\rm end}}\right) - \frac{1}{12}\ln [g_*(T_{\rm reh})] \,,
\ee
\end{widetext}
where $\rho_{\rm end}$ is the energy density at the end of the inflation, $w\approx(n-1)/(n+1)$ is the barotropic parameter during the oscillations [cf. Eq.~\eqref{w6}], $\rho_{\rm reh}$ is the energy density when the reheating takes place, and $g_*(T_{\rm reh})$ is the number of effective relativistic degrees of freedom at the reheating temperature [cf. Eq.~\eqref{Treh}]. By solving this equation numerically we can determine the correct $N_{\ast}$ for a given reheating temperature $T_{\rm reh}$ and $A_s$. In the end, $\alpha$ remains the only free parameter of the model.

\begin{figure}[ht]
    \centering
    \includegraphics[width=\textwidth]{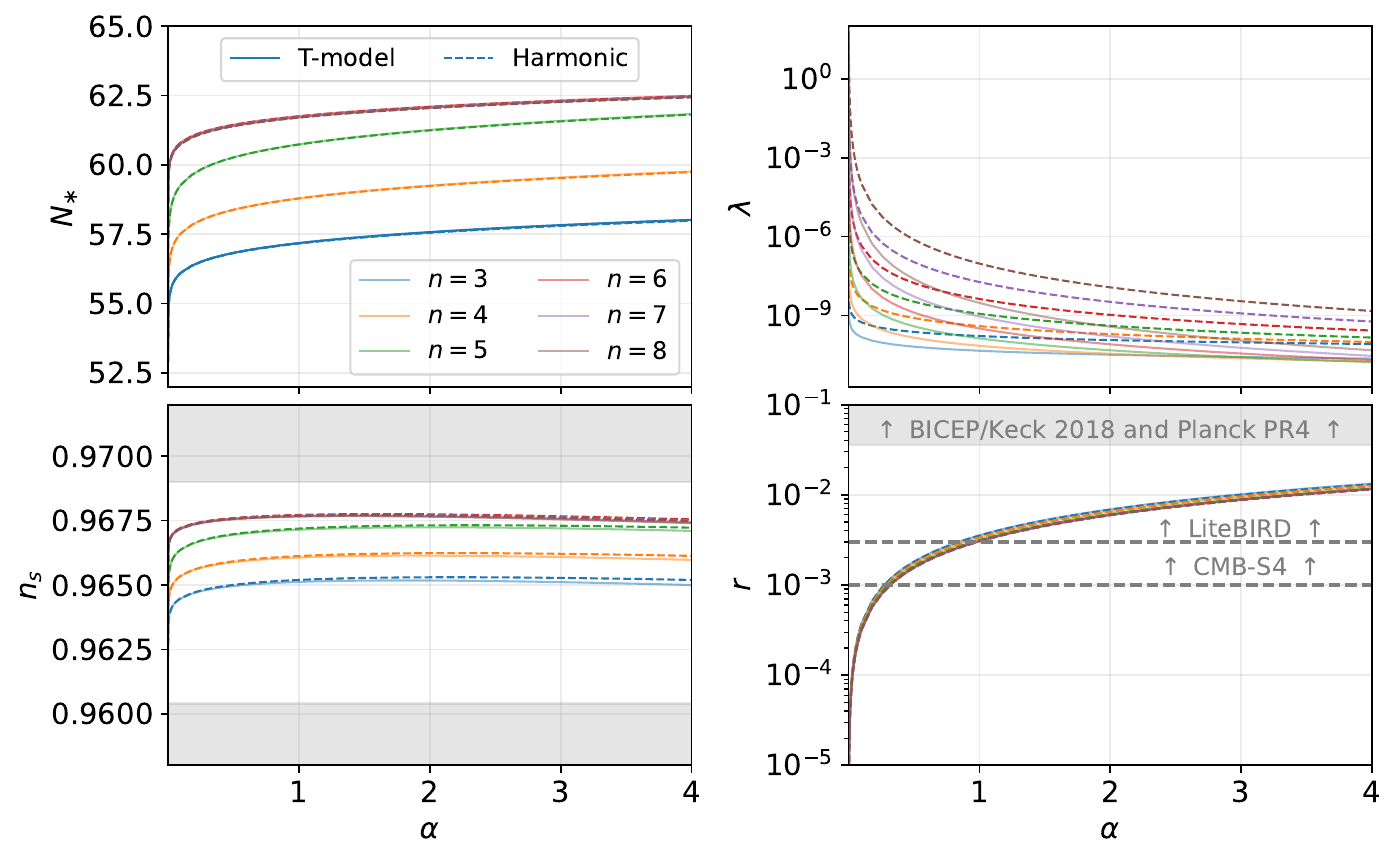}
    \caption{The plot of the inflationary observables as a function of $\alpha$ for the T-model and harmonic potentials, Eqs.~\eqref{Vvarphi6} and \eqref{VALP},     respectively, with various values of $n$. For $n=3-5$, we fix the reheating temperature $T_{\rm reh}$ to its lowest possible value using Eq. \eqref{Treh}, which ensures that the inflaton field remains homogeneous until reheating. For $n=6-8$, we set the reheating temperature to its lowest value such that the GW spectrum saturates the BBN bound for $\alpha={\cal O}(1)$. See Secs.~\ref{sec:frag} and \ref{ssec: gw-energy-spectrum} for more details on how we fix the reheating temperature. The gray regions on the spectral index plot is excluded by the Planck 2018 data~\cite{Planck:2018vyg}, while the gray region on the tensor-to-scalar ratio plot refers to the bound $r<0.03$. On the latter plot, we also show prospective reaches of the future experiments LiteBIRD~\cite{LiteBIRD:2022cnt} and CMB-S4~\cite{CMB-S4:2016ple}. We observe that LiteBIRD and CMB-S4 will be able to put constraints on this model which read approximately $\alpha \lesssim 1$ and $\alpha \lesssim 0.3$, respectively.
    }
    \label{fig:slow-roll}
\end{figure}

\begin{figure}[ht]
    \centering
    \includegraphics[width=0.8\textwidth]{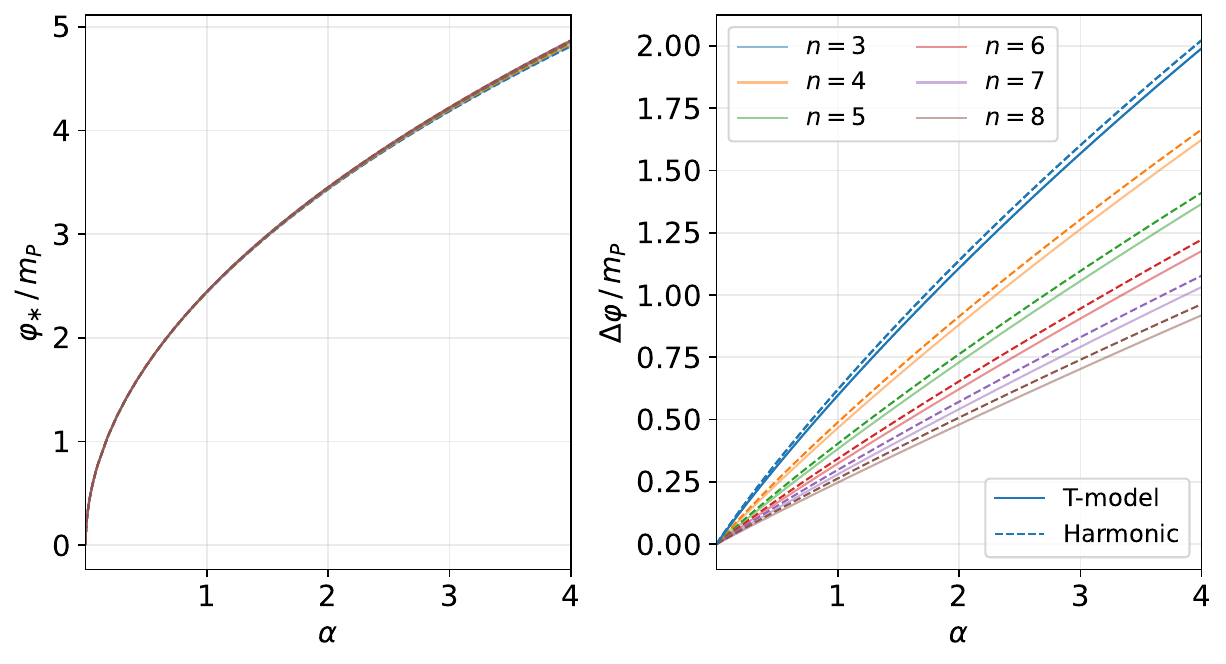}
    \caption{Left: plot showing the value $\varphi$ of the noncanonical inflaton field at horizon exit of the CMB pivot scale setting $k_p = 0.05\,$Mpc$^{-1}$ for various $\alpha$-attractor models. Right: plot showing the noncanonical inflaton field $\varphi$ excursion during inflation for various $\alpha$-attractor models. We note the values are Planckian for $\alpha={\cal O}(1)$.}
    \label{fig:field-excursion}
\end{figure}

We show our results for the inflationary observables in Fig.~\ref{fig:slow-roll}. For $n=3-5$, we fix the reheating $T_{\rm reh}$ via Eq. \eqref{Treh}, which ensures that the inflaton field remains homogeneous until reheating. See Sec. \ref{sec:frag} for the explanation. For $n=6-8$, we set the reheating temperature such that the gravitational wave spectrum saturates the BBN bound for $\alpha={\cal O}(1)$, as explained further in Sec. \ref{ssec: gw-energy-spectrum}.  We see that both potentials, monomial or sinusoidal, yield to very similar results. The upper bound $r<0.03$ on the tensor-to-scalar ratio gives an upper bound for $\alpha$ which reads \mbox{$\alpha\lesssim 10$}. As shown in Fig.~\ref{fig:slow-roll}, future experiments such as LiteBIRD~\cite{LiteBIRD:2022cnt} and CMB-S4~\cite{CMB-S4:2016ple} will improve this bound approximately to \mbox{$\alpha\lesssim 1$} and \mbox{$\alpha\lesssim 0.3$}, respectively.

The $\alpha$ parameter is also bounded by the fact that we want to avoid trans-Planckian field excursions, see Fig.~\ref{fig:field-excursion}. Thus, it is more realistic to consider \mbox{$\alpha={\cal O}(1)$}. There is no lower bound on $\alpha$ coming from the inflationary observables.

\section{Fragmentation of the oscillating condensate}
\label{sec:frag}

So far, our setup appears very promising. We may end up with a stiff phase after the end of inflation, whose barotropic parameter can be as low as \mbox{$w=1/2$} (when $n=3$), so the corresponding peak of primordial gravitational radiation is rather mild. Such a stiff phase could be prolonged without disturbing BBN, enhancing thus GWs at observable frequencies, provided reheating is not very efficient and $T_{\rm reh}$ is low. However, it turns out we cannot have too small $T_{\rm reh}$ and the whole mechanism for boosting GWs at observable frequencies is undermined as a result. The reason is the following.

A major concern is the possible fragmentation of the oscillating condensate by parametric resonant effects. This has been studied in detail in Refs.~\cite{Lozanov:2016hid,Lozanov:2017hjm,Garcia:2023dyf,Garcia:2024zir}, where it was shown that a prolonged stiff period was possible only for \mbox{$\alpha={\cal O}(1)$}. The duration of the stiff period after inflation before the fragmentation of the condensate is~\cite{Lozanov:2017hjm}
\begin{equation}
    \Delta N=\frac{n+1}{3}\ln\left(\frac{1}{\delta\, d^2}
    \frac{M}{m_P}\frac{2|2-n|}{n+1}\right),
    \label{DN1}
\end{equation}
where $d$ is the strength of the resonance band and \mbox{$\delta=0.126$} is a numerical coefficient obtained in the simulations.
The values of \mbox{$d=d(n)$} is shown in Fig.~4 of Ref.~\cite{Lozanov:2017hjm}. 

Now, provided that the condensate is not yet fragmented, the energy density of the oscillating scalar field is given by \mbox{$\rho_\phi\propto a^{-3(1+w)}$}, where \mbox{$w=w(n)$} is determined in Eq.~\eqref{w6}.
Then, the duration of the stiff period of coherent oscillations is determined by reheating. The total number of stiff $e$-folds is
\begin{eqnarray}
 &&   \Delta N=\ln\left(\frac{a_{\rm reh}}{a_{\rm end}}\right)=\frac{1}{3(1+w)}\ln\left(\frac{\rho_{\rm end}}{\rho_{\rm reh}}\right)\nonumber\\
 & \Rightarrow & \Delta N\simeq\frac{2(n+1)}{3n}\ln\left(\frac{V_{\rm end}^{1/4}}{T_{\rm reh}}\right),
    \label{DN2}
\end{eqnarray}
where we used Eqs.~\eqref{w6} and \eqref{rhoreh}, with the approximation \mbox{$(\pi^2g_*/30)^{1/4}\simeq{\cal O}(1)$}.

Therefore, the smallest possible reheating temperature corresponds to when the condensate is about to be fragmented, and it is given by
\begin{equation}
    T_{\rm reh}=\exp\left[-\frac{3n}{2(n+1)}\Delta N\right]V_{\rm end}^{1/4}\,,
    \label{Treh}
\end{equation}
where $\Delta N$ is given by Eq.~\eqref{DN1}, and $V_{\rm end}^{1/4}$ is calculated using $\phi_{\rm end}$ determined from the condition $\varepsilon(\phi_{\rm end})=1$, as shown in Fig.~\ref{fig:phiend}.

Employing Fig.~4 of Ref.~\cite{Lozanov:2017hjm} and Eqs.~\eqref{DN1} and \eqref{Treh}, we obtain the values of the reheating temperature, shown in Table~\ref{Table}, where we have assumed the maximum value \mbox{$\alpha\simeq 4$}.

\begin{widetext}
\begin{center}
\begin{table}[h]
    \centering
    \begin{tabular}{c|c|c|c|c|c|c|c|c|c}
$n$ & 2 & 3 & 4 & 5 & 6 & 7 & 8 & 9 & 10\\\hline
$d$ & 0.072 & 0.060 & 0.044 & 0.030 & 0.025 & 0.020 & 0.015 & 0.012 & 0.010\\
$\Delta N$ & 0 & 11.458 & 16.141 & 21.347 & 26.068 & 31.221 & 37.880 & 42.808 & 46.567
\\$\phi_{\rm end}/m_p$
& 2.38 & 3.17 & 3.78 & 4.27 & 4.68 & 5.03 & 5.34 & 5.62 & 5.87
\\
$T_{\rm reh}$(GeV) & 
$4.6 \times 10^{15}$ & $1.2 \times 10^{10}$ & $1.7\times 10^{7}$ & $1.1\times 10^{4}$ & $11.5$ & $6.5\times 10^{-3}$ & $2.8\times 10^{-7}$ & $3.1\times 10^{-10}$ & $6.5\times 10^{-14}$
    \end{tabular}
    \caption{The values of the strength of the resonance $d$, the number of e-dols of the corresponding stiff era $\Delta N$ and the lowest possible reheating temperature $T_{\rm reh}$, which guarantees that reheating occurs not after the fragmentation of the oscillating condensate, for a range of values of $n$. We have considered the harmonic potential, and the maximum value \mbox{$\alpha = 4$}, while the results are very similar for the T-model potential. Note that the $T_{\rm reh}$ values for $n>7$ imply that the condensate is not fragmented until reheating, which has to happen before BBN, i.e. $T_{\rm reh}\gtrsim 1\,$MeV.}
    \label{Table}
\end{table}
\end{center}
\end{widetext}

As an example, we can consider \mbox{$n=5$}, such that \mbox{$d=0.03$}, \mbox{$\Delta N \simeq 21$} and \mbox{$T_{\rm reh} \simeq 1.1\times 10^4\,$GeV}.
The corresponding harmonic scalar potential would be \mbox{$V=V_0[1-\cos(\varphi/M)]^5$}, with \mbox{$M \simeq 4.90\,m_P$} and \mbox{$V_0^{1/4} \simeq 2.6\times 10^{16}\,$GeV}, were \mbox{$\alpha=4$} and we have used Eqs.~\eqref{Malpha} and~\eqref{V0}, respectively.
If we consider the T-model potential instead, we have $V\propto\varphi^{10}$, as in Ref.~\cite{Lazarides:1986rt}. In this case, the upper bound on the quadratic mass term not to become important until reheating given in Eq.~\eqref{mTreh}, becomes \mbox{$m<0.14\,T_{\rm reh}^{8/5}/m_P^{3/5} \simeq  1.6 \times 10^{-24}\, m_P\sim 3.8 \times 10^{-6}\,$GeV}. This is rather strong which suggests that the choice of harmonic potential is more realistic.

\section{Primordial Gravitational Waves}
\label{sec:gw}

\subsection{Notations}

This part is devoted to the basic notations and dynamics of GWs. Readers who are familiar with these can skip this part.
The GWs are defined as the transverse-traceless (TT) part of the metric perturbations,
\be
\ddd s^2 = a^2(\tau) [-\ddd\tau^2 + (\delta_{ij} + h_{ij}) \ddd x^i \ddd x^j] ~,
\ee
where $\tau$ is the conformal time, and the TT gauge is given by $\delta^{ij} h_{ij} = 0$ and $\partial_i h^{ij} = 0$. The Kronecker $\delta$ symbol is used to raise/lower the spatial indices. The equation of motion for GWs is derived from the linear Einstein field equation,
\be \label{eq:eom_tensor}
h^{s''}_{\mathbf{k}}(\tau) 
+ 2 {a'\over a} h^{s'}_{\mathbf{k}}(\tau) 
+ k^2 h^s_{\mathbf{k}}(\tau) = 0 ~,
\ee
where the prime here denotes the $\tau$ derivative, and $h^s_{\mathbf{k}}$ is the Fourier mode of $h_{ij}$,
\be
h_{ij}(\tau,\mathbf{x})
= 
\sum_{s = +,\times} \int { \mathrm{d}^3 \mathbf{k} \over (2 \pi)^{3/2} } 
e^{i \mathbf{k} \cdot \mathbf{x}} e^s_{ij}(\mathbf{k}) h^s_{\mathbf{k}}(\tau) ~,
\ee
where $s = +, \times$ denote two polarizations of the GWs. The polarization tensor $e^\lambda_{ij}(\mathbf{k})$ can be expressed in terms of a pair of polarization vectors $e_i(\mathbf{k})$ and $\bar{e}_i(\mathbf{k})$, both of which are orthogonal to the wave vector $\mathbf{k}$,
\bl\label{Polarization_Tensor}
e^+_{ij}(\mathbf{k}) 
=& 
\frac{1}{\sqrt{2}} 
\l[ e_i(\mathbf{k}) e_j(\mathbf{k}) 
- \bar{e}_i(\mathbf{k}) \bar{e}_j(\mathbf{k}) \r], \nn
\\
e^\times_{ij}(\mathbf{k})
=& 
\frac{1}{\sqrt{2}} 
\l[ e_i(\mathbf{k}) \bar{e}_j(\mathbf{k}) 
+ \bar{e}_i(\mathbf{k}) e_j(\mathbf{k}) \r] ,
\el
where the unit polarized tensors satisfy the following properties: $\delta^{ij} e_{ij}^{s} = 0$, $k^i e_{ij}^{s} =0$, $\delta^{ik} \delta^{jl} e_{ij}^{s} e_{kl}^{m} = \delta^{sm}$. 

We are interested in the statistics of GWs in observation. The dimensionless power spectrum of GWs $\mathcal{P}_h(k, \tau)$ is defined through its two-point correlation function,
\be
\langle h^s_{\mathbf{k}}(\tau) h^m_{\mathbf{k}'}(\tau) \rangle 
= 
\delta^{s m} \delta^{(3)}(\mathbf{k} + \mathbf{k}') \frac{2 \pi^2}{k^3} \mathcal{P}_h(k, \tau) ~,
\ee
where we have assumed the isotropy of the stochastic GW background (SGWB), namely, $\mathcal{P}_h(k, \tau)$ depends only on the magnitude $k \equiv |\mathbf{k}|$. For simplicity, we also assume that the SGWB is unpolarized, $\langle |h^{+}_{k}| \rangle = \langle |h^{\times}_{k}| \rangle \equiv \langle |h_{k}| \rangle$. It is straightforward to show that
\be
\langle h^{ij}(\tau,\mathbf{x}) h_{ij}(\tau,\mathbf{x}) \rangle
= 2 \int {\ddd k \over k} \mathcal{P}_h(k, \tau) ~,
\ee
where $\mathcal{P}_h(k, \tau) = {k^3 \over 2 \pi^2} \langle |h_{k}(\tau)|^2 \rangle$ and $\langle \cdots \rangle$ refers to the ensemble average. For GW observations, we are concerned with the total power spectrum $2 \mathcal{P}_h(k, \tau)$. At the end of slow-roll inflation, the power spectrum for GWs is calculated as \cite{Gouttenoire:2021jhk}
\be \label{eq:ph_inf}
\mathcal{P}_h^{\rm inf}(k) \approx \frac{2}{\pi^2} \l( {H_{\inf} \over m_P} \r)^2 \l( {k \over k_p} \r)^{n_T} ~,
\ee
where the tensor spectral index $n_T$ satisfies the consistency relation $n_T \approx - 2 \epsilon \approx -r/8$, which is small in our model. The numerical result at the end of inflation is shown in Figs.~\ref{fig:gwfig} and \ref{fig:gwfig+}. 

\subsection{GW energy spectrum with the stiff period}
\label{ssec: gw-energy-spectrum}

During the vanilla slow-roll inflation (as in our model), the tensor perturbations (i.e. GWs) are frozen on the superhorizon scales and the power spectrum is nearly scale invariant. 
After inflation, the Hubble horizon grows faster than the redshift of the GWs' wavelengths ($H^{-1} \propto a^{3(1+w)/2} > a$, for $w\geq 0$ after inflation), each mode of GWs reenters the horizon at the different times and starts oscillating. These reentered modes become the part of the SGWB that we are able, in principle, to observe. In order to quantify the ability of GW detection, it is customary to define the GW density parameter on the subhorizon scales per logarithmic momentum interval \cite{Gouttenoire:2021jhk}, 
\be \label{eq:omegaGW_def}
\Omega_{\rm GW}(\tau, k) \equiv {1\over\rho_c(t)} {\ddd \rho_{\rm GW} \over \ddd \ln k} \approx {k^2 \over 12 a^2 H^2} \mathcal{P}_h(\tau,k) ~, 
\ee
where $\rho_c(\tau) = 3 m_P^2 H^2(\tau) \propto a^{-3 (1 + w) }$ is the critical density of the Universe at time $\tau$ and $k$ is related to the observed GW frequency as
\be \label{eq:freq_today}
f = {k \over 2\pi a_0} = {H_k \over 2\pi} {a_k \over a_0} ~,
\ee
where $a_k$ is the scale factor when the $k$ mode reenters the horizon.

Instead of analytically solving Eq.~\eqref{eq:eom_tensor}, it is helpful to make some reasonable estimate on the behaviors of the subhorizon GWs. It is clearly seen that the GWs' amplitudes are damped on the subhorizon scales, namely, \mbox{$h_k(\tau) \propto k^{-3/2} a_{k}/a(\tau)$}, where $k^{-3/2}$ is due to the fact that the GW power spectrum is nearly scale invariant [cf. Eq.~\eqref{eq:ph_inf}] and we dropped the Hubble friction term. Using the relationship $k = a_k H_k \propto a^{-(1+3w)/2}$, namely, replacing $a_k$ by $k^{-2/(1+3w)}$, we derive  \cite{Gouttenoire:2021jhk,Haque:2021dha,Figueroa:2019paj}\footnote{The spectrum of GWs for modes that reenter the horizon during a stiff period was first considered in Ref.~\cite{Giovannini:1998bp}.}
\be
\Omega_{\rm GW}(f) \propto f^\beta ~,
\quad {\rm where} \quad
\beta = 2\, {w - 1/3 \over w + 1/3} ~.
\label{fbeta}
\ee
Hence, for modes that reenter the Hubble horizon during RD, $w = 1/3$, the observed GW energy spectrum is flat, while for modes that reenter the horizon during the stiff period in our model $w\approx\frac{n-1}{n+1}$ [cf. Eq.~\eqref{w6}] and thus $\beta \approx \frac{2(n-2)}{2n-1}$, it generates a blue-tilted spectrum for $n>2$.\footnote{The GW spectrum in the case when the Universe is dominated by a scalar field condensate oscillating in a potential of the form $V\propto\phi^{2n}$ has been also considered in Refs.~\cite{Chakraborty:2023ocr,Barman:2024slw}, but the constraints from the possible fragmentation of the condensate were not taken into account.} For the extremely low-frequency GWs whose modes reenter the Hubble horizon during MD ($w=0$), its energy spectrum is red tilted. Simply, we can parametrize the observed GW energy spectrum consisting of the following three parts:
\be \label{eq:omegaGW_1}
\Omega_{\rm GW}(\tau_0, f) \simeq \Omega_{\rm GW}^{\rm RD}
\l\{
\begin{matrix}
    &(f/f_{\rm reh})^{\mbox{\small $\frac{2(n-2)}{2n-1}$}} ~, \quad &&f_{\rm reh} < f < f_{\rm end} ~,
    \\ \\
    &1 ~,  &&f_{\rm eq} < f < f_{\rm reh} ~,
    \\ \\
    &(f_{\rm eq}/f)^2 ~,  &&f_{0} < f < f_{\rm eq} ~,
\end{matrix}
\r. ~,
\ee
where $\Omega_{\rm GW}^{\rm RD}$ is a constant representing the GW density parameter of modes that reenter the Hubble horizon during RD. The observed frequencies $f_{\rm end}$, $f_{\rm reh}$, $f_{\rm eq}$, and $f_{0}$, correspond to the GW modes that reenter the Hubble horizon at the end of inflation, the onset of RD, the radiation-matter equality, and the present Hubble horizon, respectively. It is important to understand that there is a maximum frequency $f_{\rm end}$ because there is a minimum length scale, that is the one which exits the horizon at the end of inflation and reenters the horizon right away.
\footnote{\label{footnote7} Obviously, the assumptions made regarding the GW production near the end of inflation are not valid, as the slow roll of the inflaton field is about to be violated and there is not much time for GW states to be squeezed when exiting the horizon. This means that, very near $f_{\rm end}$, the GW spectrum deviates from Eq.~\eqref{eq:omegaGW_1}.}

In order to pin down the unknown parameters in Eq.~\eqref{eq:omegaGW_1}, we define a transfer function following Ref. \cite{Figueroa:2019paj}, 
\be
\mathcal{P}_h(\tau,k) \equiv T_h(\tau,k) \mathcal{P}_h^{\rm inf}(k) ~,
\quad
T_h(\tau,k) \equiv {1\over2} \l( {a_k \over a(\tau)} \r)^2 ~,
\ee
to quantify the time evolution between the horizon reentry $k = a_k(\tau_k) H_k(\tau_k)$ and a later time $\tau > \tau_k$. Note that the factor $1/2$ comes from the time average of the oscillating amplitudes of GWs inside the Hubble horizon, and the damping is described by $( a_k / a(\tau) )^2$. 
First, let us estimate the plateau value $\Omega_{\rm GW}^{\rm RD}$ using Eqs. \eqref{eq:ph_inf} (with $n_T=0$) and \eqref{eq:omegaGW_def} \cite{Figueroa:2019paj},
\begin{widetext}
\be
\Omega_{\rm GW}^{\rm RD} \approx {k^2 \over 12 a^2 H^2} T_h(\tau_0, k) \mathcal{P}_h^{\rm inf}(k)
\approx {1 \over 12\pi^2} \l( {g_*^k \over g_*^0} \r) \l( {g_s^0 \over g_s^k} \r)^{4/3} \Omega_{\rm rad}(\tau_0) \l( {H_{\inf} \over m_P} \r)^2
\approx 3 \times 10^{-17}~,
\ee
\end{widetext}
where we also used $H_{\inf} \approx 10^{-5} m_P$, $\Omega_{\rm rad}(\tau_0) \approx 9 \times 10^{-5}$, $g_*^k \approx g_s^k \approx 106.37$, $g_*^0 \approx 3.36$ and $g_s^0 \approx 3.91$. 

Then, we can estimate the typical frequencies of Eq. \eqref{eq:omegaGW_1} based on the relation \eqref{eq:freq_today}. The lowest frequency of SGWB is estimated as
\be
f_0 = {H_{\rm 0} \over 2\pi}
\sim 1.6 \times 10^{-43} ~{\rm GeV}
\sim 2.4 \times 10^{-19} ~{\rm Hz} ~,
\ee
where $H_0 \sim 10^{-33} ~{\rm eV}$. Similarly,
\be
f_{\rm eq} 
= {H_{\rm eq} \over 2\pi} {a_{\rm eq} \over a_0}
\sim {\sqrt{\rho_{\rm eq}} \over 2\pi \sqrt{3}\, m_P} {a_{\rm eq} \over a_0}
\sim 10^{-41} ~{\rm GeV}
\sim 10^{-17} ~{\rm Hz} ~,
\ee
where we ignored the dark energy domination and considered $T_{\rm eq} \sim 1 ~{\rm eV}$.

In order to calculate the highest frequency $f_{\rm end}$, we need to know the energy evolution after the end of inflation and until the onset of RD: the stiff period. The total energy density during the stiff period is given by $\rho_\phi \propto a^{-3(1+w)}=a^{-6n/(n+1)}$, where we used Eq.~\eqref{w6}. We thus have $\rho_{\rm end}=\rho_{\rm reh} (a_{\rm reh} / a_{\rm end})^{6n/(n+1)}$. Equations.~\eqref{Vinf} and \eqref{V0}, suggest that \mbox{$\rho_{\rm end}\sim\alpha\times 10^{-10}\,m_P^4$}, while \mbox{$\rho_{\rm reh}=\frac{\pi^2}{30} g_* T_{\rm reh}^4$}.
Thus, we can estimate
\begin{eqnarray}
  & & 
  \frac{\rho_{\rm end}}{\rho_{\rm reh}}\simeq\left(\frac{a_{\rm reh}}{a_{\rm end}}\right)^{6n/(n+1)}\sim\alpha\left(\frac{10^{16}\,{\rm GeV}}{T_{\rm reh}}\right)^4  
  \nonumber\\
  \Rightarrow & &   
  \frac{a_{\rm reh}}{a_{\rm end}}\sim\alpha^{(n+1)/6n} \left(\frac{10^{16}\,{\rm GeV}}{T_{\rm reh}}\right)^{2(n+1)/3n}\,. 
  \label{aratio}
\end{eqnarray}
Hence, the radiation energy density at the end of inflation can be estimated as 
\begin{equation}
    \rho_r^{\rm end}\simeq\rho_{\rm reh} \left(\frac{a_{\rm reh}}{a_{\rm end}}\right)^4\sim
    T_{\rm reh}^4\,
    \alpha^{2(n+1)/3n}\left(\frac{10^{16}\,{\rm GeV}}{T_{\rm reh}}\right)^{8(n+1)/3n}.
\end{equation}
With the above preparations, we calculate
\be
{a_{\rm end} \over a_0} 
\simeq {T_0 \over T_{\rm end}} 
\sim {T_{\rm CMB} \over (\rho_r^{\rm end})^{1/4}}
\sim 
10^{-29}\alpha^{-(n+1)/6n}
\left(\frac{10^{16}\,{\rm GeV}}{T_{\rm reh}}\right)^{(n-2)/3n}\,,
\label{aratio+}
\ee
where we have used that \mbox{$H_{\rm end}\sim\sqrt\alpha\times 10^{-5}\,m_P$} and the temperature of the CMB today is \mbox{$T_{\rm CMB} \sim 10^{-13} ~{\rm GeV}$}. Hence, we readily obtain
\be
f_{\rm end}=\frac{H_{\rm end}}{2\pi}\frac{a_{\rm end}}{a_0}
\sim 
\alpha^{(2n-1)/6n}
\left(\frac{10^{16}\,{\rm GeV}}{T_{\rm reh}}\right)^{(n-2)/3n}\times 10^7\,{\rm Hz}\,,
\label{fend}
\ee
where we used that \mbox{1~GeV$\, \simeq 1.5\times 10^{24}\,$Hz}. In view of Eq.~\eqref{aratio} and also considering that \mbox{$f\propto aH$}, we find
\be
f_{\rm reh} 
= f_{\rm end} {H_{\rm reh} a_{\rm reh} \over H_{\rm end} a_{\rm end}}
\simeq f_{\rm end} \l( { a_{\rm reh} \over a_{\rm end} } \r)^{(1-2n)/(n+1)}
\sim 
\left(\frac{10^{16}\,{\rm GeV}}{T_{\rm reh}}\right)^{-1}
\times 10^7\,{\rm Hz}\,,
\label{freh}
\ee
where we used that \mbox{$a\propto H^{-\frac{2}{3(1+w)}}\Rightarrow H\propto a^{-3n/(n+1)}$} and we employed Eq.~\eqref{w6} again.
From Eq.~\eqref{freh}, it is evident that the dependence of $f_{\rm reh}$ on both $\alpha$ and $n$ cancels out as it should, because these parameters influence only the physics before reheating.

With the above preparations, the current GW spectra in Eq.~\eqref{eq:omegaGW_1} are determined by parameter $\alpha$, the power index $n$, and reheating temperature $T_{\rm reh}$. 
As an example, we consider $n=5$ and $\alpha=4$, in which case Table~\ref{Table} suggests that the lowest possible reheating temperature (i.e., the longest possible stiff period) is \mbox{$T_{\rm reh} \simeq 1.1 \times 10^4\,$GeV}. Then Eq.~\eqref{fend} suggests \mbox{$f_{\rm end}\simeq 3.7 \times 10^9\,$Hz}, while Eq.~\eqref{freh} gives \mbox{$f_{\rm reh}\simeq 1.1 \times 10^{-5} \,$Hz}, with the GW spectrum growing as $\propto f^{2/3}$ in the high-frequency domain [cf. Eqs.~\eqref{fbeta} and \eqref{eq:omegaGW_1}].
\footnote{With $n=5$, in the case of monomial potential, the lower bound in Eq.~\eqref{alphabound} suggests \mbox{$\alpha>0.01$}, which is well satisfied in this example.}

\begin{figure}[ht]
    \centering
    \includegraphics[width=0.6\textwidth]{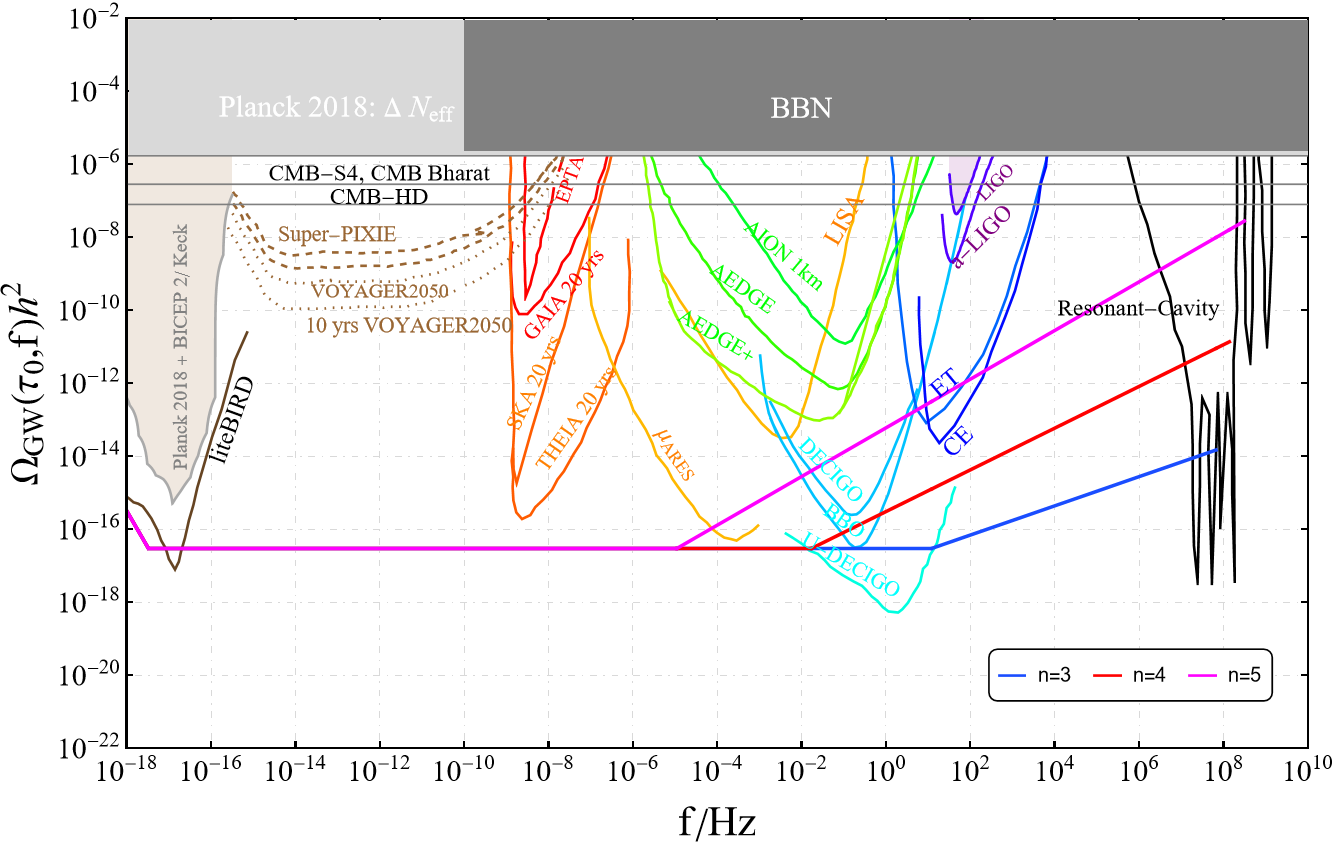}
    \caption{The current GW energy spectra in terms of various power indices of the sinusoidal potential \eqref{VALP} with $\alpha =4$: $n = 3-5$; with the corresponding reheating temperatures as shown in Table~\ref{Table}: $T_{\rm reh} = 1.2 \times 10^{10}, 1.7\times 10^{7}, 1.1\times 10^{4} ~{\rm GeV}$. The values of the stiff barotropic parameter can be calculated from Eq.~\eqref{w6}: $w = 1/2, 3/5, 2/3$. The expected sensitivity curves of various operating and forthcoming GW observatories are also shown, including ground-based interferometer detectors: LIGO/VIRGO \cite{LIGOScientific:2016aoc,LIGOScientific:2016sjg,LIGOScientific:2017bnn,LIGOScientific:2017vox,LIGOScientific:2017ycc,LIGOScientific:2017vwq}, aLIGO/aVIRGO \cite{LIGOScientific:2014pky,VIRGO:2014yos,LIGOScientific:2019lzm}, AION \cite{Badurina:2021rgt,Graham:2016plp,Graham:2017pmn,Badurina:2019hst}, Einstein Telescope (ET) \cite{Punturo:2010zz,Hild:2010id}, Cosmic Explorer (CE) \cite{LIGOScientific:2016wof,Reitze:2019iox}; space-based interferometer detectors: LISA \cite{LISA:2017pwj,Baker:2019nia}, BBO \cite{Crowder:2005nr,Corbin:2005ny}, DECIGO/U-DECIGO \cite{Seto:2001qf,Kudoh:2005as,Yagi:2011wg,Kawamura:2020pcg}, AEDGE \cite{AEDGE:2019nxb}, $\mu$-ARES, VOYAGER2050 \cite{Sesana:2019vho}; CMB spectral distortions, PIXIE/Super-PIXIE \cite{Kogut:2019vqh}; recasts of star surveys: GAIA/THEIA \cite{Garcia-Bellido:2021zgu}; CMB polarization measurements, Planck 2018 \cite{Akrami:2018odb} and BICEP 2/ Keck \cite{BICEP2:2018kqh, Clarke:2020bil}, LiteBIRD \cite{Hazumi:2019lys}, square-kilometer array (SKA) \cite{Carilli:2004nx,Janssen:2014dka,Weltman:2018zrl}, EPTA \cite{Lentati:2015qwp,Babak:2015lua}, NANOGRAV \cite{McLaughlin:2013ira,NANOGRAV:2018hou,Aggarwal:2018mgp,Brazier:2019mmu,NANOGrav:2020bcs,NANOGrav:2023hvm,NANOGrav:2023gor}; conversion into electromagnetic waves, the resonant cavity experiments~\cite{Herman:2022fau,Aggarwal:2020olq,Ringwald:2022xif}
    }
    \label{fig:gwfig}
\end{figure}

The current GW energy spectra determined by Eq. \eqref{eq:omegaGW_1} are shown in Fig.~\ref{fig:gwfig} for $n=3-5$ and Fig.~\ref{fig:gwfig+} for $n=6-8$, respectively, along with various operating and forthcoming GW experiments summarized in the caption of Fig.~\ref{fig:gwfig}. In both figures, we take the maximum value $\alpha = 4$. In Fig.~\ref{fig:gwfig}, we choose the lowest possible reheating temperatures as shown in Table~\ref{Table}: $T_{\rm reh} = 1.2 \times 10^{10}$, $1.7\times 10^{7}$, and $1.1\times 10^{4} ~{\rm GeV}$ for $n=3,4,$ and $5$, respectively, corresponding to possible largest peaks in GW spectra, implied by Eqs.~\eqref{eq:omegaGW_1} and  \eqref{freh}.
It is clearly seen from Fig.~\ref{fig:gwfig} that their GW spectra excess sensitivity curves of several GW experiments. In particular, they are all detectable by the resonant cavity experiments in the high-frequency range $10^{6}-10^{9}$ Hz, which targets electromagnetic signals generated by GWs in resonant cavity experiments~\cite{Herman:2022fau,Aggarwal:2020olq}.
\footnote{Several promising bounds on the GWs in the high-frequency domain have been proposed recently~\cite{Domcke:2022rgu, Bringmann:2023gba, Liu:2023mll, Ito:2023fcr, Ito:2023nkq,Gatti:2024mde} (also see review~\cite{Aggarwal:2020olq}), which, however, are weaker than BBN constraint by orders of magnitude currently.}

The ultrahigh-frequency GW (UHF-GW) initiative~\cite{Aggarwal:2020olq}, which has recently come into the picture, discusses several prospects for detecting very high-frequency GWs leading to new ideas for detection techniques, e.g., \cite{Ballantini:2005am,Arvanitaki:2012cn,Ejlli:2019bqj,Aggarwal:2020umq,LSD:2022mpz,Berlin:2021txa,Berlin:2022hfx,Berlin:2023grv,Goryachev:2021zzn,Goryachev:2014yra,Campbell:2023qbf,Sorge:2023nax,Tobar:2023ksi,Carney:2023nzz,Domcke:2022rgu,Domcke:2023bat,Bringmann:2023gba,Vacalis:2023gdz,Liu:2023mll,Ito:2019wcb,Ito:2020wxi,Ito:2022rxn,Ito:2023bnu}.
Still, it remains extremely difficult experimentally, to go beyond the BBN bound.
With our analysis on primordial GWs, which originated during inflation, we aim to motivate future investigation and provide a concrete science case for UHF-GW detectors. It is remarkable to have the possibility to probe beyond the SM microphysics physics at energy scales many orders of magnitude beyond the reach of conventional cosmological tools at our disposal or laboratory experiments. Particularly, in Figs.~\ref{fig:gwfig} and \ref{fig:gwfig+}, we showed that resonant cavity experiments \cite{Herman:2022fau} could potentially observe primordial GW backgrounds that nearly saturate the upper bound.
\footnote{It also maybe possible that the SM contribution to SGWB from thermal fluctuations may contribute significantly the total gravitational wave background in those regions, see  \cite{Ghiglieri:2015nfa,Ghiglieri:2020mhm,Drewes:2023oxg}.}

\begin{figure}[ht]
    \centering
    \includegraphics[width=0.4\textwidth]{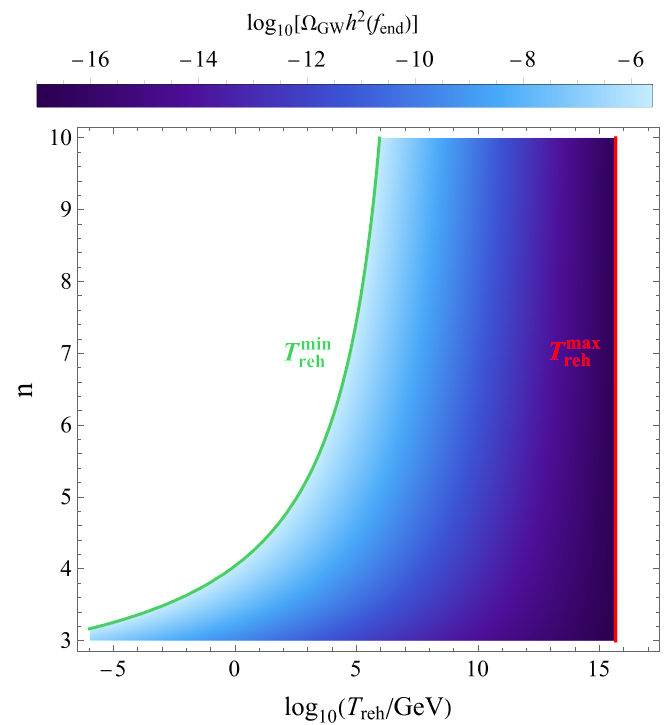}
    \caption{The allowed parameter space $(n, T_{\rm reh} )$ by considering the BBN bound $\Omega_{\rm GW} h^2 \leq 2.2 \times 10^{-6}$, while the color refers to the corresponding peak value $\Omega_{\rm GW} h^2(f_{\rm end})$.
    The BBN bound provides more stringent lower bound $T_{\rm reh}^{\rm min}$ (the green curve) compared to the fragmentation effect shown in Table~\ref{Table}. The upper bound is $T_{\rm reh}^{\rm max} \simeq 4.6 \times 10^{15}~{\rm GeV}$ for $\alpha =4$ from the fact that $\rho_{\rm reh} \leq \rho_{\rm end}$.
}
    \label{fig:region}
\end{figure}

\begin{figure}[ht]
    \centering
    \includegraphics[width=0.6\textwidth]{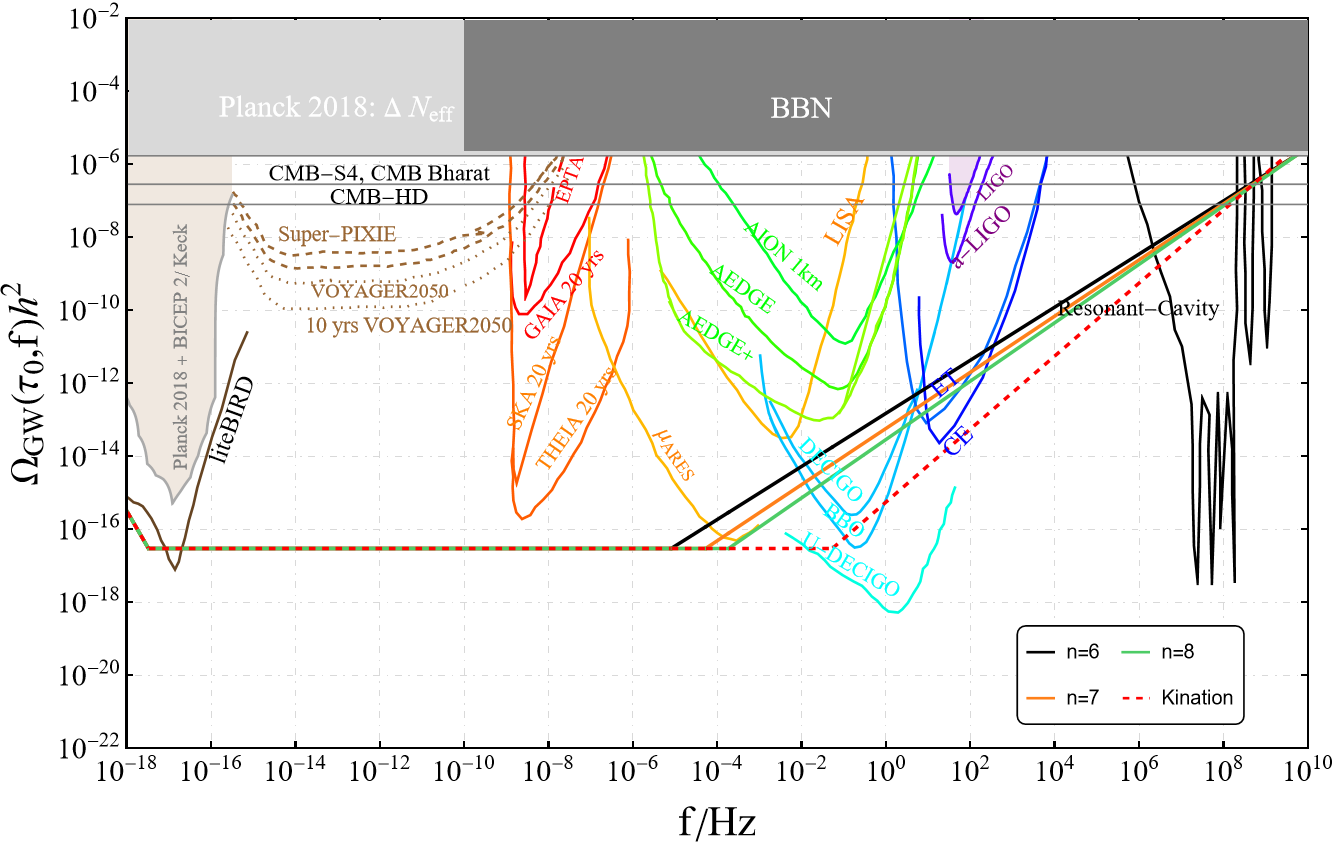}
    \caption{The GW energy spectra in terms of various power indices of the polynomial potential in Eq.~\eqref{Vvarphi6} with $\alpha = 4$, $n = 6, 7, 8$; with the corresponding reheating temperatures, $T_{\rm reh} = 8.1 \times 10^3, 5.3 \times 10^4, 1.8 \times 10^5 ~{\rm GeV}$, such that the peaks of the GW spectra saturate the BBN bound $\Omega_{\rm GW} h^2 \leq 2.2 \times 10^{-6}$ at the highest frequencies $f_{\rm end} \simeq 7.2 \times 10^9~{\rm Hz}$.     
    The values  of the stiff barotropic parameter can be calculated from Eq.~\eqref{w6}, $w = 5/7, 3/4, 7/9$. 
    The red dashed curve refers to the kination proper with $w=1$ ($n\rightarrow\infty$), $T_{\rm reh} \simeq 5.3 \times 10^{7}~{\rm GeV}$, $f_{\rm reh} \simeq 5.3 \times 10^{-2}~{\rm Hz}$. It is clear that the produced GW spectrum for small $n$ (but $n>5$) is significantly boosted compared to the case when regular kination follows inflation, as is usual in many quintessential inflation models (see e.g., Ref.~\cite{Dimopoulos:2022rdp} and references therein).
    The expected sensitivity curves of various operating and forthcoming GW observatories are the same as in Fig.~\ref{fig:gwfig}.
    }
    \label{fig:gwfig+}
\end{figure}

For $n>5$, the lower bounds on reheating temperatures from the fragmentation effect are weaker (as shown in Table~\ref{Table}) than the BBN bound
\footnote{CMB bound on $\Delta N_{\rm eff}$ is comparable to BBN; see, for instance, \cite{Ben-Dayan:2019gll}.} $\Omega_{\rm GW} h^2 \leq 2.2 \times 10^{-6}$, which is able to provide more stringent lower bounds $T_{\rm reh}^{\rm min}$. This is shown by the green curve in Fig.~\ref{fig:region}, which shows the allowed values of $T_{\rm reh}$ for different values of $n$, given the BBN bound and $\alpha = 4$.
\footnote{Releasing $\alpha$ helps provide more information on the allowed values of $T_{\rm reh}$ for different $n$, since $\alpha$ affects the peak of $\Omega_{\rm GW}$ only through $f_{\rm end}$ [c.f. Eq.~\eqref{fend}], and there is a very weak dependence on $n$ in the power of $\alpha$ in Eq.~\eqref{fend}, except for a extremely tiny $\alpha$. Our calculations have confirmed this argument, so we only show the two-parameter region [c.f. Fig.~\ref{fig:region}], instead of the full three-parameter region.} 
In addition, the upper bound on reheating temperature comes from the fact that $\rho_{\rm reh} \leq \rho_{\rm end}$, which gives $T_{\rm reh}^{\rm max} \simeq 4.6 \times 10^{15}~{\rm GeV}$ for $\alpha = 4$ (which is consistent with the case $n=2$ in Table~\ref{Table}), as shown by the red vertical line in Fig.~\ref{fig:region}. The color in Fig.~\ref{fig:region} denotes the peak value $\Omega_{\rm GW} h^2(f_{\rm end})$ for a given set $(n, T_{\rm reh})$. It is straightforward to see that, for a fixed $n$, the peak value becomes smaller as the reheating temperature becomes higher. This is expected, because the $f_{\rm reh}$ shifts to higher values.
Moreover, the lower bound $T_{\rm reh}^{\rm min}$ tends to a constant $\simeq 5.3 \times 10^{7}~{\rm GeV}$ for large $n$, which is suggested by Eqs.~\eqref{eq:omegaGW_1} and \eqref{fend} in the large-$n$ limit, namely the growth rate of GW spectra becomes $\propto f$ and $f_{\rm end} \rightarrow  \alpha^{1/3} ( 10^{16}\,{\rm GeV} / T_{\rm reh} )^{1/3} \times 10^7\,{\rm Hz}$. Hence, the peak of GW spectra in the large-$n$ limit is given by $\Omega_{\rm GW}(f_{\rm end}) \sim \Omega_{\rm GW}^{\rm RD} \alpha^{1/3} ( 10^{16}\,{\rm GeV} / T_{\rm reh} )^{4/3}$. The corresponding lowest possible value of $f_{\rm reh}$ for a large $n$ is calculated as $f_{\rm reh} \sim 5.3 \times 10^{-2}~{\rm Hz}$. Thus, it is hard to detect their GW spectra with the operating and forthcoming GW experiments shown in Fig.~\ref{fig:gwfig}, expect for the resonant cavity. All the above findings in the large-$n$ limit are reasonable, since the potential in Eqs.~\eqref{Vvarphi6} or~\eqref{VALP} approximates a square potential well after inflation, such that the inflaton field takes less time to reach the potential's minimum and reheating would roughly happen afterward.

In Fig.~\ref{fig:gwfig+}, the lowest bounds $T_{\rm reh} = 8.1 \times 10^3$, $5.3 \times 10^4$, and $1.8 \times 10^5 ~{\rm GeV}$ are taken for $n=6,7,$ and $8$ respectively, such that the GW spectra saturate the BBN bound $\Omega_{\rm GW} h^2 \leq 2.2 \times 10^{-6}$ at a nearly identical highest frequency $f_{\rm end} \simeq 7.2 \times 10^9~{\rm Hz}$ (which also shows the weak dependence on $n$ in $f_{\rm end}$). As shown in Fig.~\ref{fig:gwfig+}, the cases $n=6-8$ are also detectable by the resonant cavity at the high-frequency band and several GW experiments at the low-frequency band including the Einstein Telescope.
\footnote{It is important to remark that during our analysis we chose the spectral tilt $n_T$ of the tensor modes that excited during inflation to be negligibly small. This is true when the primordial inflation is well described by a quasi-de Sitter background, which is assumed in this work. However, several alternative scenarios exist where $n_T$ could be very different and the inflationary tensor spectrum consists of a large blue tilt for the modes that excited during the inflationary epoch or even during the postinflationary era, for instance, due to particle production. Nonetheless, we refrain from considering $n_T$ to be a free parameter, as this would have also given us another independent variable to chose during inflation. In order to see the constraints on GW signals when assuming that $n_T$ is a free parameter, see e.g., Refs.~\cite{Zhao:2011bg,Zhao:2013bba,Kuroyanagi:2014nba,Jinno:2014qka,Lentati:2015qwp,Lasky:2015lej,Arzoumanian:2015liz,Liu:2015psa,Kuroyanagi:2018csn,DEramo:2019tit,Bernal:2019lpc}. For concrete theory realizations for blue-tilted $n_T \ge 0$  see e.g., string gas cosmology \cite{Brandenberger:2006xi}, superinflation models \cite{Baldi:2005gk}, G inflation \cite{Kobayashi:2010cm}, noncommutative inflation \cite{Calcagni:2004as,Calcagni:2013lya}, particle production during inflation \cite{Cook:2011hg,Mukohyama:2014gba}, and several others~\cite{Kuroyanagi:2020sfw}.}

\section{The astrophysical foreground}

For the stochastic GW of cosmological origin, one may expect many astrophysical sources of GWs. LIGO/VIRGO has already observed binary black hole (BH-BH) \cite{LIGOScientific:2016vbw, LIGOScientific:2016sjg, LIGOScientific:2016dsl} as well as binary neutron star (NS-NS) \cite{LIGOScientific:2017vwq} merging events.
In order to distinguish the SGWB sourced by inflationary tensor perturbations with stiff pre-BBN era and those from the one generated by the astrophysical foreground, one should expect the NS and BH foreground might be subtracted with sensitivities of BBO and ET or CE windows, possibly during the range $\Omega_{\rm GW} \sim 10^{-15}$ \cite{Cutler:2005qq} and $\Omega_{\rm GW} \sim 10^{-13}$ \cite{Regimbau:2016ike}. The binary white dwarf galactic and extra-galactic foreground could dominate over the NS-NS and BH-BH foregrounds in LISA~\cite{Farmer:2003pa, Rosado:2011kv, Moore:2014lga}, and should be subtracted~\cite{Kosenko:1998mv} with the expected sensitivity $\Omega_{\rm GW} \sim 10^{-13}$ to be reached at LISA~\cite{Adams:2010vc, Adams:2013qma}. Given such subtractions could be made possible in the future along with the crucial fact that the GW spectrum generated by the astrophysical foreground increases with frequency as $f^{2/3}$ \cite{Zhu:2012xw}, that is, completely different from the GW spectrum inflationary gravitational waves in the stiff period $f^{2(\frac{n-2}{2n-1})}$ (unless $n=5$), as suggested by Eq.~\eqref{eq:omegaGW_1}, one may envisage to pin down the GW signals from inflationary first-order tensor perturbation. Moreover, our mechanism clearly overwhelms the astrophysical GW background at high frequencies, when $n>5$.

\section{Discussion and Conclusion}
\label{sec:conclusion}

Various cosmological sources such as strong first-order phase transitions, cosmic strings or domain walls, inflationary preheating, etc. lead to detectable GWs of stochastic origin from the early Universe, which provides a unique opportunity to peek into the pre-BBN epoch. Particularly, this is useful in probing new physics beyond the SM, as for example, GUT-scale physics, high scale physics related to dark matter physics, and matter-antimatter asymmetry \cite{Dasgupta:2022isg,Bhaumik:2022pil,Barman:2022yos,Ghoshal:2022jdt,Dunsky:2021tih,Bernal:2020ywq,Ghoshal:2020vud}, which are otherwise beyond the reach of LHC or any other laboratory or astrophysical searches for new physics due to heavy scales involved.

Most compellingly, inflation generically gives rise to a stochastic GW background, which extends to very high frequencies, determined by the inflationary energy scale. However, the conventional inflationary paradigm generates primordial GWs, which are too faint to be observable in the near future. Fortunately, such weak GWs can be boosted to observability if inflation is followed by a period with a stiff equation of state. Depending on the barotropic parameter $w$ of the stiff phase, the GW spectrum features a peak toward large frequencies. If the peak is too sharp, then BBN considerations do not allow the boosted spectrum to extend to observable frequencies. What is needed is a stiff period with $\frac13<w<1$, which lasts for a long time corresponding to late reheating \cite{Figueroa:2019paj}.
\footnote{Also see Ref.~\cite{SanchezLopez:2023ixx}.}

In this work, we proposed two concrete inflationary scenarios which naturally lead to nonstandard cosmological evolution (with an appropriate, stiff equation of state in the postinflationary era) with potentially large GW signals arising from first-order inflationary tensor perturbations. We then study how tensor perturbations generated during inflation may be amplified during the stiff era and investigate whether they lead to detectable signals for gravitational waves detectors such as LISA, ET, u-DECIGO, and BBO. 

In particular, we consider a scalar field condensate, oscillating in a potential well of the form \mbox{$V(\varphi)\propto\varphi^{2n}$}. The oscillating scalar condensate is characterized on average by a barotropic parameter \mbox{$w=(n-1)/(n+1)$}, which can take a value inside the range \mbox{$\frac13<w<1$}, when \mbox{$n\geq 3$}. If the Universe is dominated by our oscillating scalar field, then it would engage in a stiff period as desired. Before the oscillations, our scalar field can be the inflaton, because its kinetic term features a pole, following the $\alpha$-attractors construction.
\footnote{Such poles in the kinetic term may also give rise to primordial black hole formation and scalar-induced GWs \cite{Ghoshal:2023pcx,Afzal:2024xci}.}
\footnote{It should be noted that one would not need to rely on $\alpha$-attractors for the formation of the inflationary plateau. Other proposals would lead to very similar results, e.g., shaft inflation \cite{Dimopoulos:2014boa}.}

Even though our theoretical framework allows us to obtain a stiff period with multiple values of the barotropic parameter inside the desired window, our setup does not allow a very low reheating temperature $T_{\rm reh}$ when the order $n$ of the scalar potential is small. This is because the oscillating condensate tends to fragment due to resonance effects. Thus, if this is the case, reheating must occur before this fragmentation takes place if we want to remain in the stiff period before reheating. This would correspond to a lower bound on the reheating temperature. As a result, when $3\leq n\leq 5$ the peak attained in the GW spectrum cannot be extended to very low frequencies. Nevertheless, contact with the forthcoming observations can indeed be achieved, especially for $n=5$, as shown in Fig.~\ref{fig:gwfig}.

When $n>5$, the lower bound on $T_{\rm reh}$ is very weak and the actual constraint on the boosted GW spectrum is due to the requirement that the GW peak does not challenge the process of big bang nucleosynthesis (and also Planck-CMB) \mbox{$\Omega_{\rm GW}(f_{\rm end})<10^{-6}$}. As a result, the GW spectrum can indeed be extended down to observable frequencies, especially for $n=6-8$ or even higher. Indeed, as shown in our Fig.~\ref{fig:gwfig+}, there is clear overlap with the projected observations of DECIGO, $\mu$ARES, Big Bang Observatory, Cosmic Explorer, and Einstein Telescope. Moreover, in Fig.~\ref{fig:gwfig+}, it is demonstrated that the GW spectrum is clearly enhanced compared to the case of kination proper (with $w=1$) following inflation, as usual in quintessential inflation models. The maximum enhancement is achieved with $n=6$. 

As shown in Figs.~\ref{fig:gwfig} and \ref{fig:gwfig+}, we  manage to obtain a characteristic GW spectrum, boosted from the scale-invariant vanilla case. If future observations do detect such kind of spectrum of primordial GWs, then we will obtain crucial information of the physics of inflation \cite{Mishra:2021wkm}, such as the inflation energy scale or the value of the reheating temperature, as well as the steepness (the order $2n$) of the scalar potential near its minimum. 

An important point to stress has to do with the sharp peak in the GW spectrum at frequency $f_{\rm end}$, corresponding to the end of inflation. The GW spectrum in Eq.~\eqref{eq:omegaGW_1} was obtained under the assumption of slow roll. However, as explained in footnote~\ref{footnote7}, we expect the slow-roll approximation to cease to be valid near the end of inflation, when the $\epsilon$ slow-roll parameter is no longer much smaller than unity, so $\epsilon<1$ but not $\ll 1$. This happens about an $e$-fold before the end of inflation. Now, a sizable $\epsilon$ means that $H$ is not robust anymore, because \mbox{$\epsilon=-\dot H/H^2$}. Because the GW spectrum is proportional to $H^2$ [cf. Eq.~\eqref{eq:ph_inf}], this really means that near the peak the spectrum is somewhat suppressed (the peak is rounded) as $H$ begins to decrease. This suggests that the BBN and the CMB $N_{\rm eff}$ bounds in Figs.~\ref{fig:gwfig} and \ref{fig:gwfig+} are less challenged, i.e., that our results are even safer.

It is intriguing that due to the existence of a nonstandard postinflationary pre-BBN cosmology plays the morphology of the gravitational wave spectrum for a given microscopic physics scenario is characterized. Furthermore, this postinflationary epoch may also leave signatures in the CMB spectrum itself as it may impact the number of $e$-folds during inflation, thereby correlating predictions for the inflationary observables such as $n_s$ and $r$ with GW signals for a given inflationary scenario.

\medskip
\section*{Acknowledgement}
C.C. thanks Jie Jiang, Chen Zhang for useful discussions. 
C.C. is supported in part by the National Key R\&D Program of China (Grant No. 2021YFC2203100). K.D. is supported (in part) by the STFC consolidated Grant: ST/X000621/1. C.E. received support from the 2236 Co-Funded Brain Circulation Scheme2 (CoCirculation2)
of the Scientific and Technological Research Council of Türkiye T\"UB\.{I}TAK (Project No 121C404).

\medskip

\bibliographystyle{apsrev4-1}
\bibliography{kination2}

\end{document}